\documentclass[12pt]{article}
\pdfoutput =1
\usepackage{float} 
\usepackage{amsmath}
\usepackage{slashed}
\usepackage{amsfonts}   
\usepackage{amssymb}

\begin{document}
\date{}
\title{{\bf{\Large Stringy correlations on deformed $ AdS_{3}\times S^{3} $}}}
\author{
 {\bf {\normalsize Dibakar Roychowdhury}$
$\thanks{E-mail:  dibakarphys@gmail.com, dibarak@post.bgu.ac.il}}\\
 {\normalsize  Department of Physics, Ben-Gurion University of The Negev,}\\
  {\normalsize P.O. Box 653, Beer-Sheva 84105, Israel}
\\[0.3cm]
}

\maketitle
\begin{abstract}
In this paper, following the basic prescriptions of Gauge/String duality, we perform a strong coupling computation on \textit{classical} two point correlation between \textit{local} (single trace) operators in a gauge theory dual to $ \kappa $-deformed $ AdS_{3}\times S^{3}$ background. Our construction is based on the prescription that relates every local operator in a gauge theory to that with the (semi)classical string states propagating within the \textit{physical} region surrounded by the holographic screen in deformed $ AdS_3 $. In our analysis, we treat strings as being that of a point like object located near the physical boundary of the $ \kappa $- deformed Euclidean Poincare $ AdS_{3} $ and as an extended object with non trivial dynamics associated to $ S^{3} $. It turns out that in the presence of small background deformations, the usual power law behavior associated with two point functions is suppressed exponentially by a non trivial factor which indicates a faster decay of two point correlations with larger separations. On the other hand, in the limit of large background deformations ($ \kappa \gg 1 $), the corresponding two point function reaches a point of saturation. In our analysis, we also compute finite size corrections associated with these two point functions at strong coupling. As a consistency check of our analysis, we find perfect agreement between our results to that with the earlier observations made in the context of vanishing deformation.
\end{abstract}

\section{Overview and Motivation}
The classic mathematical evidences regarding the existence of an integrable structure on both sides of the $ AdS_{5}/CFT_{4} $ duality \cite{Maldacena:1997re} might be regarded as one of the major theoretical advancements that took place during the past one and half decade \cite{Serban:2010sr}-\cite{Ryang:2006yq}. It turns out that, in the so called planar limit, the dilatation operator associated with $ \mathcal{N}=4 $ SYM could be mapped to that with the corresponding Hamiltonian of an integrable spin chain in one dimension \cite{Minahan:2002ve}. On the other hand, the integrable structure associated with the stringy side of the duality has been ensured due to the existence of an infinite umber of conserved quantities associated with the Lagrangian field equations in $ AdS_{5}\times S^{5}$ \cite{Bena:2003wd}. 

During the past one and half decades, the quest for an integrable deformation corresponding to $ AdS_{5}\times S^{5} $ superstring theory has been one of the prime focus of modern theoretical investigation \cite{Lunin:2005jy}-\cite{Matsumoto:2014ubv}. Very recently, the novel discovery \cite{Delduc:2013qra} regarding the one parameter integrable deformation associated with $ AdS_{5}\times S^{5} $ superstring sigma model has drawn renewed attention due to its several remarkable features. At this stage one should take a note on the fact that the deformed sigma model \cite{Delduc:2013qra} had been formulated in the presence of a real deformation parameter ($ \eta $) such that the model exhibits two characteristic features-(1) the presence of Lax connection and (2) the invariance under the kappa symmetry. The kappa symmetry associated with the deformed superstring model turns out to be absolutely essential in order to ensure a type IIB supergarvity background.

Soon after the discovery of this new class of integrable deformations \cite{Delduc:2013qra}, the corresponding deformed target spacetime metric was figured out by authors in \cite{Arutyunov:2013ega} where, considering the so called light cone gauge, they perform the perturbative ($ 2 \rightarrow 2 $) $ S $ matrix computation in the Hamiltonian framework. It is also noteworthy to mention that in their analysis the authors restricted themselves only to the bosonic sector of the full theory. On of the key outcomes of their analysis was the fact that in the limit of the large string tension, the $ S $ matrix corresponding to the integrable $ q $- deformed model was found to be in a perfect agreement to that with the corresponding perturbative $ S $ matrix computed for the $ \eta $ deformed model once various other parameters of the $ q $- deformed theory could be related to that with the real deformation parameter $ \eta \in [0,1) $ in a following manner,
\begin{eqnarray}
\kappa =\frac{2\eta}{1-\eta^{2}},~~q=e^{-\frac{\vartheta}{T}},~~\vartheta =\frac{2\eta}{1+\eta^{2}}.\label{Kappa}
\end{eqnarray}

The computation \cite{Arutyunov:2013ega} further unveils the fact that the full $ 10D $ background corresponding to the $ NS-NS $ sector supports a metric together with some non vanishing $ B $ field. The metric contribution in the bosonic sector of the Lagrangian could be divided into two pieces namely, the deformed $ AdS_{5} $ and the deformed $ S^{5} $. The  Wess-Zumino sector of the bosonic Lagrangian, on the other hand, sources the non trivial $ B $ field in the target spacetime. Given the above relation (\ref{Kappa}), there are several interesting limits that one might wish to explore. For example, the limit $ \eta \rightarrow 0$ clearly reproduces the undeformed $ AdS_5 \times S^{5} $ background. On the other hand, in the limit, $ \eta \rightarrow 1$ the original $ AdS_5 \times S^{5} $ gets mapped into $ dS_5 \times H^{5} $ indicating the fact that the corresponding world sheet theory is non unitary \cite{Hoare:2014pna}. Therefore the deformation acts as an interpolation between $ AdS_5 \times S^{5} $ and $ dS_5 \times H^{5} $. In our analysis, while solving the corresponding stringy dynamics associated to deformed $ AdS_{3}\subset AdS_{5} $, we focus particularly in these two limits in order to gain further insights regarding the interpolating regime. 

Before we actually explain the purpose of our present analysis, it worth emphasizing that the deformed model proposed in \cite{Delduc:2013qra}-\cite{Delduc:2014kha} leaves behind it many open issues that need to be addressed properly. Here we elaborate some of them. Due to the presence of the curvature singularity at some finite radial distance, $ \varrho \sim \kappa^{-1} $ it turns out that strings are eventually confined within a region, $ 0<\varrho< \kappa^{-1}$. The vanishing of the beta function \cite{Appadu:2015nfa} somehow guarantees that such deformations might be allowed by string theory although its implication is not very much clear at this moment.  As a consequence of this, the deformed target space metric corresponding to $ AdS_{5} $ appears to be with no boundary in the usual sense \cite{Hoare:2014pna}. Instead one could think of a \textit{holographic screen}  \cite{Kameyama:2014vma}-\cite{Kameyama:2014via} and solve stringy dynamics within the region bounded by this holographic screen. It turns out that the region bounded by this holographic screen is the only allowed physical region in the bulk where the classical string solutions as well as the holographic correspondence make sense \cite{Kameyama:2014vma}-\cite{Kameyama:2014via}. The notion of the usual boundary ($ \varrho \rightarrow \infty $) could however be recovered only in the limit of the vanishing deformation. As a matter of fact, it turns out that the spacetime supersymmetry associated with the target spacetime is lost and on top of it the bosonic isometry associated with the undeformed $ 10D $ background gets $ q $-deformed and/or hidden to its smaller subset. In other words, the original $ SO(2,4)\times SO(6) $ isometry is found to be broken down to its Cartan subgroup $ U(1)^{3}\times  U(1)^{3} $ \cite{Hoare:2014pna} corresponding to shifts along various bosonic directions. As a natural consequence of this, the corresponding interpretation and/or the implication of these broken symmetry generators on the properties of the dual gauge theory is not clear immediately. In other words, it is not known a priori how various gauge invariant operators and in particular their correlation functions would be modified under this reduced subset of symmetry generators. However, the symmetries associated with the deformed $10D  $ background immediately suggests that the dual gauge theory should not manifest any conformal invariance as well as supersymmetry. Keeping these facts in mind, it seems quite urgent to build up the necessary mathematical framework that would eventually unveil the hidden symmetries associated with this mysterious dual gauge theory at least in the regime of strong coupling. These are the precisely the issues that we consider to be worthy of further investigation.

 In order to address the above mentioned issues in a systematic way, in the present paper we carry out a classical computation on the correlation function \cite{Janik:2010gc}-\cite{Bozhilov:2011qf} between \textit{heavy} local operators in the gauge theory those are dual to classical spinning strings moving non-trivially over $ \kappa $- deformed background \cite{Hoare:2014pna}. It is therefore a strong coupling computation from the perspective of the dual gauge theory. In our analysis, we consider only the bosonic sector of the full superstring theory \cite{Arutyunov:2013ega}. Our analysis might be regarded as being the straightforward generalization of the earlier proposal in the context of $ AdS_{5}\times S^{5}$ superstrings \cite{Janik:2010gc}, where we compute two point correlations between single trace operators those are dual to \textit{semi-classical} string states propagating over $ \kappa $- deformed  $ AdS_{5}\times S^{5}$ geometry \cite{Hoare:2014pna}. In our analysis, we consider two types of operators in the dual gauge theory namely, the magnons and the spikes.
 
According to the methods developed in \cite{Janik:2010gc}, the Polyakov action (corresponding to these semi-classical string states over $ \kappa $- deformed geometry) evaluated at the classical saddle point should correspond to the desired two point correlation between the single trace operators in the dual gauge theory at strong coupling. For the case of usual $ AdS_{5}\times S^{5}$ superstrings, the same prescription provides the correct two point correlation between local operators in the gauge theory where one could easily identify the corresponding classical conformal dimension to that with the energy of the stringy excitation in the bulk \cite{Park:2010vs}. However, this is not the scenario that one should even expect to be hold true when the original bosonic isometry associated with the target spacetime is broken. Instead, it is quite natural to expect that the usual power law behavior \cite{Park:2010vs} associated with these two point correlations would be modified in a non trivial fashion.
 
Before we actually start describing the precise mathematical framework adopted in our computation, it is customary to mention that instead of considering the full $ 10D $ background, we perform our analysis over the truncated target space of the full deformed geometry namely, the $ \kappa $- deformed $ AdS_{3}\times S^{3} $ \cite{Hoare:2014pna}. As long as one is concerned only with the bosonic $ NS-NS $ sector \cite{Arutyunov:2013ega} of the the full superstring theory, the resulting metric corresponding to this truncated model turns out to be the direct sum of the two individual sectors namely the $ AdS_{3} $ and the $ S^{3} $. This picture no longer holds true as soon as one turns on $ RR $ fields\footnote{In \cite{Lunin:2014tsa}, the authors had explored the corresponding $ 6D $ supergravity background supported by $ \kappa $- deformed $ AdS_{3}\times S^{3} $. In their analysis, the authors had constructed one parameter family of supergravity solution (in the presence of non trivial $ RR $ fields, dilation and the three form) in $ 6D $ as a consistent truncation of type $ IIB$ $ 10D $ supergravity on $ T^{4} $. However, it turns out that for the supergravity background to be a consistent solution, the parameter cannot be arbitrary and has to take certain specific values in different limiting conditions.}  \cite{Lunin:2014tsa}.  This $ 6D $ analogue of the full $ 10D $ background possesses two basic characteristic features namely, (1) the integrability of this model is ensured from the very outset and (2) the vanishing of the corresponding $ B $ field that was originally present in $ 10D $ \cite{Hoare:2014pna}. Moreover, it turns out that this $ \kappa $- deformed $ AdS_{3}\times S^{3} $ model admits a RG flow \cite{Hoare:2014pna} (in the same sense as that of the two parameter deformed $ O(4) $ sigma model) where corresponding to a large value of the deformation parameter ($ \kappa \rightarrow \infty $), the theory flows to a UV fixed point namely, $ dS_{3}\times H^{3} $.

In our analysis, we consider two types of operators in the dual gauge theory namely the magnons and the spikes \cite{Park:2010vs} and we compute the associated two point correlations for each of these operators separately. As far as the bulk picture is concerned, we treat strings as being that of a point like object located near the \textit{physical} boundary (the so called holographic screen \cite{Kameyama:2014vma}-\cite{Kameyama:2014via}) of the $ \kappa $- deformed Euclidean Poincare $ AdS_{3} $ and as an extended object moving non trivially over three sphere ($ S^{3} $). From our analysis, it turns out that stringy fluctuations corresponding to deformed $ S^{3} $ are \textit{exactly} solvable in the presence of generic background deformations ($ \kappa $). However, it turns out to be extremely difficult to solve the same for generic $ \kappa $- deformations associated to $ AdS_{3} $. Therefore, in our analysis, we choose to explore the corresponding stringy excitation (associated to deformed $ AdS_{3} $) both in the perturbative ($ 0<\kappa \le 1 $) as well as in the non perturbative ($ \kappa\gg 1 $) regime. Our analysis, therefore clearly reveals an intuitive picture regarding the behavior of the two point function corresponding to some intermediate value associated with the background deformations ($ \kappa $) during the RG flow. 

Considering the perturbative regime associated with $ AdS_3 $, the corresponding strong coupling behavior associated with the two point function takes the following form,
\begin{eqnarray}
\langle \mathcal{O}(0)\mathcal{O}(x_{f})\rangle_{\kappa}  \sim \left( \frac{\epsilon}{x_{f}}\right)^{2\Delta_{\kappa}}~  e^{-\frac{2\kappa^{2}\mathfrak{q}\Delta_{\kappa}}{\nu}\log^{2}|\frac{\epsilon}{x_{f}} |+\mathcal{O}(\kappa^{4})}\label{2}
\end{eqnarray}
which clearly reveals the fact that the usual power law fall off \cite{Park:2010vs} is exponentially suppressed by a non trivial factor. Stating in another way, the above formula (\ref{2}) provides the small deformation behavior of the two point function between local operators in a holographic RG flow. The exponential suppression above in (\ref{2}) has its origin in the stringy dynamics corresponding to deformed Euclidean $ AdS_{3} $ sector of the full theory. However, at this stage, it is worth emphasizing that the associated power law behavior (\ref{2}) (which is the reminiscent of the usual power law fall off in a CFT \cite{Park:2010vs}) is \textit{exact} (in the sense that we determine the coefficient $ \Delta_{\kappa} $ exactly in terms of the corresponding background deformation ($ \kappa $)) and has its origin in the stringy dynamics associated with deformed $ S^{3} $. Therefore, considering the above facts together, one might treat (\ref{2}) as being that of a \textit{semi-perturbative} expression for the two point function in the gauge theory.

On the other hand, the nontrivial \textit{leading} order contributions corresponding to the non perturbative regime ($ \eta \rightarrow 1$, $ \kappa \gg 1 $) has its source in the background deformations associated with the deformed $ S^{3} $ sector of the full geometry namely,
\begin{eqnarray}
\langle \mathcal{O}(0)\mathcal{O}(x_{f})\rangle_{\kappa \rightarrow \infty}\sim e^{iS^{3}} \sim   e^{-\frac{ \hat{T}\bar{s}\mathfrak{C}}{\kappa^{2}}+\mathcal{O}\left( \frac{1}{\kappa^{4}}\right)}\sim const.
\end{eqnarray}
 It turns out that the contribution from the deformed $ AdS_{3} $ sector appears only in the subleading order and which might be regarded as the consequence of the fact that the physical $ AdS_{3} $ region allowed for strings eventually shrinks to zero in the corresponding limit. In other words, at leading order the deformed $ AdS_3 $ contribution is suppressed compared to that of the contribution associated with deformed $ S^{3} $. Also it should be kept in mind that this large deformation ($ \kappa \rightarrow \infty $) regime would correspond into the enhancement of the so called \textit{unphysical} domain that has its route in the non-unitarity associated with the world sheet theory for superstrings in the curved background \cite{Hoare:2014pna}. To summarize: (1) the two point function corresponding to small background deformations exhibits a faster fall off than what is expected from the perspective of the usual CFT and (2) it gradually saturates to some constant value for large background deformations.

The organization for the rest of the analysis is the following:
In Section 2, we solve the stringy dynamics associated with the $ \kappa $- deformed $ AdS_{3}\times S^{3} $ background. We use these solutions in Section 3, in order to compute the two point correlations between \textit{giant} mognons at strong coupling.  We also compute the effects of incorporating finite size corrections \cite{Lee:2011fe} to these correlation functions at strong coupling. We perform identical analysis for spikes in Section 4. At this stage, it is also noteworthy to mention that in the limit of the vanishing deformation, all our results matches smoothly to that with the earlier findings of \cite{Park:2010vs} where one could easily identify the entity, $ \Delta_{\kappa=0}=\Delta=E_{string} $ as being that of the classical conformal dimension associated with local operators in the dual gauge theory. Finally we conclude in Section 5.

\section{Strings in deformed $ AdS_3 \times S^{3} $ }
We start our analysis by considering the Polyakov action for open strings over the $ \kappa $- deformed geometry\footnote{The classical integrability of the deformed model (\ref{E1}) is ensured from the fact that the deformed $ S^{3} $ could be identically mapped to the corresponding two parameter deformation of $ O(4) $ sigma model under special circumstances namely, with the choice of the left right symmetric deformation that correctly reproduces the symmetries associated with the $ 3D $ target spacetime \cite{Hoare:2014pna}. The latter model was proved to be equivalent to (classical) integrable Fateev's two parameter $ SU(2) $ principal chiral model \cite{Fateev:1996ea}.} \cite{Hoare:2014pna},
\begin{eqnarray}
ds^{2}=-\mathfrak{h}(\varrho)dt^{2}+\mathfrak{f}(\varrho)d\varrho^{2}+\varrho^{2}d\psi^{2}+\tilde{\mathfrak{h}}(\theta)d\varphi^{2}+\tilde{\mathfrak{f}}(\theta)d\theta^{2}+\cos^{2}\theta d\phi^{2}
\label{E1}
\end{eqnarray}
where, the individual metric coefficients could be formally expressed as,
\begin{eqnarray}
\mathfrak{h}&=&\frac{1+\varrho^{2}}{(1-\kappa^{2}\varrho^{2})},~~\mathfrak{f}=\frac{1}{(1+\varrho^{2})(1-\kappa^{2}\varrho^{2})}\nonumber\\
\tilde{\mathfrak{h}}&=&\frac{\sin^{2}\theta}{(1+\kappa^{2}\cos^{2}\theta)},~~\tilde{\mathfrak{f}}=\frac{1}{(1+\kappa^{2}\cos^{2}\theta)}
\end{eqnarray}
such that the NS-NS two form vanishes during the process of consistent reduction from $ AdS_5 \times S^{5}$ \cite{Hoare:2014pna}. Notice that, here $ \varphi $, $ \theta $ and $ \phi $ are the angular coordinates on deformed $ S^{3} $. 

From (\ref{E1}), it is therefore indeed quite evident that there is as such no mixing between the coordinates of $ AdS_{3} $ to that with the coordinates corresponding to $ S^{3} $. Hence we can analyze them separately. 

\subsection{Solutions in $ AdS_{3} $}
\subsubsection{$\kappa$- deformed Euclidean Poincare $ AdS_3 $}
Our goal in this analysis would be to solve the (point particle) dynamics associated with the Polyakov action corresponding to open string configurations over the curved background (\ref{E1}). We perform our \textit{analytic} computations with the choice of the conformal gauge conditions for the Polyakov action. In our analysis, we would treat strings as being that of a point like object located near the holographic screen \cite{Kameyama:2014vma}-\cite{Kameyama:2014via} of the $ \kappa $- deformed Euclidean Poincare $ AdS_{3} $ and thereby ignore fluctuations on the world sheet of the string. Before we proceed further, it is also noteworthy to mention that our results are perturbative as we retain ourselves only upto leading order in the background deformations.

The deformed $ AdS_3 $ sector of the full spacetime could be formally expressed as,
\begin{eqnarray}
ds^{2}_{AdS_3}=-\mathfrak{h}(\varrho) dt^{2}+\mathfrak{f}(\varrho)d\varrho^{2}+\varrho^{2}d\psi^{2}\label{E3}
\end{eqnarray}
where we implement the following change of variables,
\begin{eqnarray}
\varrho = \sinh \chi
\end{eqnarray}
which finally yields,
\begin{eqnarray}
ds^{2}_{AdS_3}=-\left( \frac{\cosh^{2}\chi}{1-\kappa^{2}\sinh^{2}\chi}\right)dt^{2}+\frac{d\chi^{2}}{(1-\kappa^{2}\sinh^{2}\chi)}+\sinh^{2}\chi d\psi^{2}.\label{E5}
\end{eqnarray}

Clearly, the above metric (\ref{E5}) is expressed in the so called \textit{global} coordinates. However, for the sake of our present calculation, we need to re-express (\ref{E5}) in the \textit{Euclidean} Poincare coordinates \cite{Kameyama:2014vma}-\cite{Kameyama:2014via}  which corresponds to Wick rotating the real time axis, $ t \rightarrow i \tilde{t} $. 

In order to proceed further, we make the following choice,
\begin{eqnarray}
\cosh \chi =\frac{1}{\cos \gamma}.\label{E6}
\end{eqnarray}

Using (\ref{E6}), we finally rewrite the Euclidean $ AdS_{3} $ as,
\begin{eqnarray}
ds^{2}_{EAdS_3}=\frac{(d\tilde{t}^{2}+d\gamma^{2})}{\cos^{2}\gamma -\kappa^{2}\sin^{2}\gamma}+\frac{\sin^{2}\gamma}{\cos^{2}\gamma}d\psi^{2}.\label{E7}
\end{eqnarray}

Next, we substitute,
\begin{eqnarray}
z=e^{\tilde{t}}\cos \gamma ,~~r=e^{\tilde{t}} \sin \gamma
\label{E8}
\end{eqnarray}
into (\ref{E7}), which yields,
\begin{eqnarray}
ds^{2}_{EAdS_3}=\frac{1}{z^{2}}\left( \frac{dz^{2}+dr^{2}}{1-\frac{\kappa^{2}r^{2}}{z^{2}}} + r^{2}d\psi^{2}\right).
\end{eqnarray}

Finally, we perform another set of coordinate transformations namely,
\begin{eqnarray}
x^{0}=\mathfrak{t}=r \sin \psi ,~~x^{1}=\mathfrak{x}= r \cos \psi
\end{eqnarray}
which precisely ends up being giving rise to the so called Euclidean Poincare $ AdS_{3} $ associated with non trivial $ \kappa $- deformations,
\begin{eqnarray}
ds^{2}_{EAdS_3}=\frac{1}{z^{2}}\left( \frac{dz^{2}+dx^{2}}{1-\frac{\kappa^{2}x^{2}}{z^{2}}} \right)-\frac{\kappa^{2}(\mathfrak{t}d \mathfrak{x}-\mathfrak{x}d \mathfrak{t})^{2}}{z^{2}(1-\frac{\kappa^{2}x^{2}}{z^{2}})}
\label{E11}
\end{eqnarray}
where, we denote, $ x^{2}=(x^{0})^{2}+(x^{1})^{2}=\mathfrak{t}^{2}+\mathfrak{x}^{2} $. The above equation (\ref{E11}) corresponds to the exact form of the Euclidean Poincare metric with $ \kappa $- deformations. For the sake of our current analysis, we re-express (\ref{E11}) as,
\begin{eqnarray}
ds^{2}_{EAdS_3}=\mathcal{Z}(z,x)dz^{2}+\mathcal{T}(z,x)d\mathfrak{t}^{2}+\mathcal{X}(z,x)d\mathfrak{x}^{2}+2\mathcal{K}(z,x)d\mathfrak{t}d\mathfrak{x}\label{E12}
\end{eqnarray}
where, the above metric functions could be formally expressed as,
\begin{eqnarray}
\mathcal{Z}(z,x)&=&\frac{1}{z^{2}(1-\frac{\kappa^{2}x^{2}}{z^{2}})},~~
\mathcal{T}(z,x)=\frac{1-\kappa^{2}\mathfrak{x}^{2}}{z^{2}\left( 1-\frac{\kappa^{2}x^{2}}{z^{2}}\right) }\nonumber\\
\mathcal{X}(z,x)&=&\frac{1-\kappa^{2}\mathfrak{t}^{2}}{z^{2}\left( 1-\frac{\kappa^{2}x^{2}}{z^{2}}\right) },~~
\mathcal{K}(z,x)=\frac{\kappa^{2}\mathfrak{t x}}{z^{2}\left( 1-\frac{\kappa^{2}x^{2}}{z^{2}}\right) }.\label{E13}
\end{eqnarray}

Before we conclude this Section, it is noteworthy to mention that the metric singularity corresponding to $ |\frac{x}{z}|=\frac{1}{\kappa} $ above in (\ref{E12}) is an artifact of the inherent curvature singularity associated with the string metric that was present originally at, $ \varrho =\kappa^{-1} $  \cite{Hoare:2014pna}. On top of it, from the structure of the above singularity, it is indeed evident that the maximum spatial volume that one could associate with the holographic screen \cite{Kameyama:2014vma}-\cite{Kameyama:2014via} cannot be arbitrary and in fact it is fixed by the corresponding location of the screen at a fixed radial distance ($ z=z_B $) in the bulk and vice verse. In other words, if $ z=z_{B} $ be the radial location of the holographic screen in the bulk, then the maximal spatial region that one could associate with this screen (for generic background deformation ($ \kappa $)) is given by, 
\begin{eqnarray}
\mathcal{A}_{B}=\frac{\pi z_B^{2}}{\kappa^{2}}\label{A}
\end{eqnarray}
such that, $ z_B\leq z<\infty $. In other words, the gauge invariant (local) operators as well as their correlation functions in the dual gauge theory are defined only within the spatial volume, $ \mathcal{A}\leq \mathcal{A}_{B} $ associated with the holographic screen. Clearly, the area associated with the holographic screen shrinks to zero ($ \mathcal{A}_{B} \rightarrow 0 $) in the limit of the large background deformations ($ \kappa \gg 1 $). On the other hand, the volume becomes infinitely large in the limit, $ \kappa \rightarrow 0 $. Finally, considering all the previous arguments it should be clear by now that the volume, $ \mathcal{A}=\mathcal{A}_{B} $ could be regarded as being that of the \textit{minimal} spatial region that one could associate with $ z=const. $ hypersurface corresponding to a given background deformation ($ \kappa $) in the bulk.

\subsubsection{Perturbative solutions}
Our next task would be to substitute (\ref{E11}) into the Polyakov action and solve the corresponding dynamics on the world sheet. In order to solve these fluctuations, we make the following ansatz corresponding to the coordinates on the world sheet \cite{Park:2010vs},
\begin{eqnarray}
z=z(\tau),~\mathfrak{t}=\mathfrak{t}(\tau),~~\mathfrak{x}=\mathfrak{x}(\tau)
\end{eqnarray}
which finally results in the Polyakov action of the following form,
\begin{eqnarray}
S_{PAdS_3}=\frac{\hat{T}}{2}\int d^{2}\sigma\left( \mathcal{Z}(z,x)(\partial_{\tau}z)^{2}+\mathcal{T}(z,x)(\partial_{\tau}\mathfrak{t})^{2}+\mathcal{X}(z,x)(\partial_{\tau}\mathfrak{x})^{2}+2\mathcal{K}(z,x)(\partial_{\tau}\mathfrak{t})(\partial_{\tau}\mathfrak{x})\right) 
\label{E15}
\end{eqnarray}
where, $ \hat{T}=T\sqrt{1+ \kappa^{2}} $ (with, $ T=\frac{\sqrt{\lambda}}{2\pi} $) is the \textit{effective} string tension \cite{Hoare:2014pna} associated with the deformed background.

In order to proceed further, we first note down the equations of motion associated with the coordinates on the world sheet namely, $ z(\tau) $ and, $ x^{a}(\tau) $ where, we combine the remaining set of variables into a single variable, $ x^{a}=\{x^{0}=\mathfrak{t},x^{1}=\mathfrak{x}\} $ which yield the following set of equations\footnote{Here, prime corresponds to derivatives w.r.t $ \tau $.} in the background deformations,
\begin{eqnarray}
z''(\tau)-\frac{z'^{2}(\tau)}{z(\tau)}+\frac{x'^{a2}}{z(\tau)}+\kappa^{2}\Gamma (\tau)+\mathcal{O}(\kappa^{4})&=&0\nonumber\\
x''^{a}(\tau)-\frac{2x'^{a}z'}{z(\tau)}+\kappa^{2}\Xi^{a} (\tau)+\mathcal{O}(\kappa^{4})&=&0
\label{E16}
\end{eqnarray}
where, the functions $ \Gamma(\tau) $ and $ \Xi^{a}(\tau) $ could be formally expressed as,
\begin{eqnarray}
\Gamma (\tau)&=&\frac{2x^{2}z'^{2}}{z^{3}}+z^{2}\left( \frac{x^{2}z'}{z^{4}}\right)' -\frac{1}{z}(\mathfrak{x}^{2}\mathfrak{t}'^{2}+\mathfrak{t}^{2}\mathfrak{x}'^{2})+\frac{2x^{2}}{z^{3}}(\mathfrak{t}'^{2}+\mathfrak{x}'^{2})+\frac{2 \mathfrak{t x}\mathfrak{t}'\mathfrak{x}'}{z}\nonumber\\
\Xi^{a} (\tau)&=& -\frac{\mathfrak{t}z'^{2}}{z^{2}}-\frac{\mathfrak{t}}{z^{2}}(\mathfrak{t}'^{2}+\mathfrak{x}'^{2})-z^{2}\left( \frac{\mathfrak{x}^{2}\mathfrak{t}'}{z^{2}}\right)'+\mathfrak{t}\mathfrak{x}'^{2}-\mathfrak{x}\mathfrak{t}'\mathfrak{x}'
+z^{2}\left(\frac{\mathfrak{t}\mathfrak{x}\mathfrak{x}'}{z^{2}} \right)'+z^{2}\left( \frac{x^{2}\mathfrak{t}'}{z^{4}}\right)'.
\end{eqnarray}

In order to solve (\ref{E16}) perturbatively in the deformation parameter ($ \kappa $), we consider the following expansion for the variables namely, 
\begin{eqnarray}
z(\tau)=z^{(0)}(\tau)+\kappa^{2}z^{(1)}(\tau)+\mathcal{O}(\kappa^{4})\nonumber\\
x^{a}(\tau)=x^{a(0)}(\tau)+\kappa^{2}x^{a(1)}(\tau)+\mathcal{O}(\kappa^{4}).\label{E18}
\end{eqnarray}

Substituting (\ref{E18}) into the equations of motion (\ref{E16}), we first note down the zeroth order equations,
\begin{eqnarray}
z''^{(0)}-\frac{z'^{(0)2}}{z^{(0)}}+\frac{x'^{a(0)2}}{z^{(0)}}&=&0\nonumber\\
x''^{a(0)}-\frac{2x'^{a(0)}z'^{(0)}}{z^{(0)}}&=&0.\label{E19}
\end{eqnarray}

In order to solve (\ref{E19}), we first note that,
\begin{eqnarray}
x'^{a(0)}=\frac{z^{2(0)}}{\sqrt{2}}.\label{E20}
\end{eqnarray}

Substituting (\ref{E20}) into (\ref{E19}), we finally solve zeroth order equations,
\begin{eqnarray}
z^{(0)}(\tau)&=&\frac{1}{\cosh \beta \tau}\nonumber\\
x^{a(0)}(\tau)&=&\frac{1}{\sqrt{2}}\tanh \beta \tau 
\end{eqnarray}
which thereby correspond to a specific parametrization of geodesics in $ AdS_{3} $ \cite{Janik:2010gc}.

Our next task would be to use these zeroth order solutions in order to obtain the leading order corrections due to the background deformations ($ \kappa $). The corresponding equations at leading order turn out to be,
\begin{eqnarray}
z''^{(1)}-\frac{z'^{(0)2}}{z^{(0)}}\left(\frac{2z'^{(1)}}{z'^{(0)}}-\frac{z^{(1)}}{z^{(0)}} \right)+\frac{x'^{a(0)2}}{z^{(0)}}\left(\frac{2x'^{a(1)}}{x'^{a(0)}}-\frac{z^{(1)}}{z^{(0)}} \right) +\Gamma^{(0)}(\tau)&=& 0\nonumber\\
x''^{a(1)}-\frac{2x'^{a(0)}z'^{(0)}}{z^{(0)}}\left(\frac{x'^{a(1)}}{x'^{a(0)}}+\frac{z'^{(1)}}{z'^{(0)}}-\frac{z^{(1)}}{z^{(0)}} \right) +\Xi^{(0)}(\tau)&=&0\label{E22}
\end{eqnarray}
where, the entities like, $ \Gamma^{(0)}(\tau) $ and $ \Xi^{(0)}(\tau) $ comprise of all the zeroth order solutions in $ \kappa $.  A straightforward computation yields the following,
\begin{eqnarray}
\Gamma^{(0)}(\tau)&=& -\beta^{2}\sinh\beta\tau \tanh\beta\tau \nonumber\\
\Xi^{(0)}(\tau)&=&\frac{\beta^{2}}{\sqrt{2}}\tanh\beta\tau.\label{E23}
\end{eqnarray}

Substituting (\ref{E23}) into (\ref{E22}), one finds,
\begin{eqnarray}
z''^{(1)}+2\beta \tanh\beta\tau z'^{(1)}-\frac{\beta^{2}}{2}(1-3 \tanh^{2}\beta\tau ) z^{(1)}+\left( \frac{\sqrt{2}\beta }{\cosh \beta\tau}\right) x'^{a(1)}-\beta^{2}\sinh\beta\tau \tanh\beta\tau =0\nonumber\\
x''^{a(1)}+2\beta \tanh\beta\tau x'^{a(1)}-\left( \frac{\sqrt{2}\beta }{\cosh \beta\tau}\right) z'^{(1)}-\sqrt{2}\beta^{2}\left( \frac{\tanh \beta\tau}{\cosh \beta\tau}\right) z^{(1)}+\frac{\beta^{2}}{\sqrt{2}}\tanh\beta\tau = 0.\label{E24}
\end{eqnarray}

The first equation in (\ref{E24}) could be schematically expressed as,
\begin{eqnarray}
(\mathcal{D}_{\tau}+\mathfrak{m}^{2}) z^{(1)}&=&\mathfrak{J}_{(z)}(\tau) \nonumber\\
\mathfrak{J}_{(z)}(\tau)&=&-\sqrt{2}\beta \cosh\beta\tau x'^{a(1)}+\beta^{2}\sinh^{2} \beta\tau \cosh\beta\tau
\end{eqnarray}
where, $ \mathcal{D}_{\tau}=\partial_{\tau}(\cosh^{2}\beta\tau \partial_{\tau}) $. One could express the corresponding solution in terms of Green's function that eventually yields,
\begin{eqnarray}
z^{(1)}(\tau)=- \int d\tau' \mathcal{G}_{(z)}(\tau , \tau')\mathfrak{J}_{(z)}(\tau')
\end{eqnarray}
subjected to the fact,
\begin{eqnarray}
(\mathcal{D}_{\tau}+\mathfrak{m}^{2})\mathcal{G}_{(z)}(\tau , \tau')=-\delta (\tau -\tau'). 
\end{eqnarray}

Let us now focus on the second equation in (\ref{E24}). Like in the previous example, this equation could also be expressed schematically as,
\begin{eqnarray}
\mathcal{D}_{\tau}x^{a(1)}&=&\mathfrak{J}_{(x)}(\tau) \nonumber\\
\mathfrak{J}_{(x)}(\tau)&=&\sqrt{2}\beta \cosh\beta\tau z'^{(1)}+\sqrt{2}\beta^{2}\sinh\beta\tau z^{(1)}-\frac{\beta^{2}}{\sqrt{2}}\sinh \beta\tau \cosh\beta\tau.
\end{eqnarray}
The solution turns out to be,
\begin{eqnarray}
x^{a(1)}(\tau)=-\int d\tau' \mathcal{G}_{(x)}(\tau , \tau')\mathfrak{J}_{(x)}(\tau')
\end{eqnarray}
where, $ \mathcal{G}_{(x)}(\tau , \tau') $ is the corresponding Green's function that satisfies,
\begin{eqnarray}
\mathcal{D}_{\tau} \mathcal{G}_{(x)}(\tau , \tau') =-\delta (\tau -\tau')
\end{eqnarray}
together with some specific boundary conditions that will be discussed below\footnote{In the Appendix, we provide the most general form of the solution corresponding to these Green's functions in the frequency ($ \mathfrak{w} $) space.}.

Therefore, the complete set of solutions upto leading order in the deformation turns out to be,
\begin{eqnarray}
z(\tau)&=& \frac{1}{\cosh \beta \tau} -\kappa^{2}\int d\tau' \mathcal{G}_{(z)}(\tau , \tau')\mathfrak{J}_{(z)}(\tau')+\mathcal{O}(\kappa^{4}) \nonumber\\
x^{a}(\tau)&=& \frac{1}{\sqrt{2}}\tanh \beta \tau -\kappa^{2}\int d\tau' \mathcal{G}_{(x)}(\tau , \tau')\mathfrak{J}_{(x)}(\tau')+\mathcal{O}(\kappa^{4}).\label{E31}
\end{eqnarray}

Substituting (\ref{E31}) into (\ref{E15}) and retaining ourselves upto leading order in the deformation ($ \kappa $) we find,
\begin{eqnarray}
S_{PAdS_3}&=&\frac{\hat{T}}{2}(\mathcal{S}^{(0)}+\kappa^{2}\mathcal{S}^{(1)})+\mathcal{O}(\kappa^{4})\nonumber\\
&=&\frac{\hat{T}}{2}\int_{-s/2}^{s/2}d\tau \int_{-L}^{L}d\sigma \beta^{2} \gamma (\tau)\nonumber\\
\gamma (\tau)&=&1+\kappa^{2}\mathfrak{K}(\tau)+\mathcal{O}(\kappa^{4})\label{E32}
\end{eqnarray}
where, the full Polyakov action could be formally expressed as the sum of the usual \textit{on-shell} piece without any deformation \cite{Janik:2010gc},
\begin{eqnarray}
\mathcal{S}^{(0)}=\int_{-s/2}^{s/2}d\tau \int_{-L}^{L}d\sigma \beta^{2}
\end{eqnarray}
and the contribution to the on-shell action sourced due to the background deformations namely,
\begin{eqnarray}
\mathcal{S}^{(1)}=\int_{-s/2}^{s/2}d\tau \int_{-L}^{L}d\sigma \beta^{2}\mathfrak{K}(\tau)
\end{eqnarray}
where, the function,
\begin{eqnarray}
\mathfrak{K}(\tau)=2\beta^{-1} \sinh\beta\tau \int d\tau' \mathcal{G}'_{(z)}(\tau , \tau')\mathfrak{J}_{(z)}(\tau')+2\sinh\beta\tau \tanh\beta\tau \int d\tau' \mathcal{G}_{(z)}(\tau , \tau')\mathfrak{J}_{(z)}(\tau')\nonumber\\
-2\sqrt{2}\beta^{-1} \int d\tau' \mathcal{G}'_{(x)}(\tau , \tau')\mathfrak{J}_{(x)}(\tau')+\frac{2}{\cosh\beta\tau}\int d\tau' \mathcal{G}_{(z)}(\tau , \tau')\mathfrak{J}_{(z)}(\tau')+\sinh^{2}\beta\tau
\end{eqnarray}
yields the first non trivial correction to the Polyakov action in $ AdS_3 $ in the presence of the background $ \kappa $- deformations.

Note that, in order to arrive at the equations of motion, we had deliberately dropped the boundary terms (associated with a constant $ \tau $ surface) in the Polyakov action which thereby invokes certain boundary conditions for the fields on the world-sheet. We are now in a position to implement these boundary conditions \cite{Janik:2010gc}. The first set of boundary conditions that we implement is the following \cite{Janik:2010gc},
\begin{eqnarray}
x^{a}(-s/2)=0\nonumber\\
z(-s/2)=\epsilon. \label{E36}
\end{eqnarray}

On the other hand, the second set of boundary conditions turn out to be \cite{Janik:2010gc},
\begin{eqnarray}
x^{a}(s/2)=x_{f}\nonumber\\
z(s/2)=\epsilon \label{E38}
\end{eqnarray}
where, $ \epsilon $ is some appropriate UV cutoff (in the presence of background deformations) such that, $ |\epsilon -z_B|\ll 1 $ together with the fact that the bound, $ x^{2}_{f} \gtrsim \mathcal{A}_{B}$ is satisfied. 

Considering the zeroth order solutions for $ \beta (\approx \frac{2}{s}\log (x_{f}/\tilde{\epsilon} )) $ (where, $ \tilde{\epsilon} (\ll 1) $ is the usual UV cutoff without background deformations \cite{Janik:2010gc}), it is in fact quite intuitive to note that to leading order in the deformations,
\begin{eqnarray}
\mathcal{G}_{(x)}(\pm s/2 , \tau')\simeq \mathcal{G}_{(z)}(\pm s/2 , \tau')\approx \frac{ \tilde{\epsilon} ^{2}}{2 \pi x^{2}_{f}}\int \frac{d\mathfrak{w}}{\mathfrak{w}^{2}}e^{-i\mathfrak{w}\beta(\pm s/2 -\tau')}\label{e40}
\end{eqnarray} 
which therefore yields a vanishing contribution at the end points of the time evolution. 

Using (\ref{e40}), it is now quite trivial to note down, 
\begin{eqnarray}
z(\pm s/2)&=& \frac{1}{\cosh \beta (\pm s/2)} +\mathcal{O}(\kappa^{2}\tilde{\epsilon}^{2}) \nonumber\\
x^{a}(\pm s/2)&=& \frac{1}{\sqrt{2}}\tanh \beta (\pm s/2) +\mathcal{O}(\kappa^{2}\tilde{\epsilon}^{2})
\end{eqnarray} 
where, we could ignore the subleading contributions as they appear as a fourth order term in the perturbative expansion. Therefore this implies that upto leading order in the deformation,
\begin{eqnarray}
\beta \simeq\frac{2}{s}\log (x_{f}/\epsilon )+..
\end{eqnarray}
\subsubsection{A note on large $ \kappa $ solutions}
In this Section, we study equations of motion corresponding to the large values of the background deformations ($ \kappa \gg 1$). In the presence of large background deformations, the corresponding Polyakov action (\ref{E15}) takes the following form,
\begin{eqnarray}
S_{PAdS_3}=\frac{\hat{T}}{2}\int  \frac{d^{2}\sigma}{x^{2}}\left( \mathfrak{x}\mathfrak{t}' -\mathfrak{t}\mathfrak{x}'\right)^{2} +\mathcal{O}\left( \frac{1}{\kappa^{2}}\right). 
\end{eqnarray}

The corresponding leading order equations of motion turn out to be,
\begin{eqnarray}
\mathfrak{t}'\left( 1+\frac{\mathfrak{t}^{2}}{x^{2}}\right)+\frac{\mathfrak{x}\mathfrak{t}\mathfrak{x}'}{x^{2}}=0 \nonumber\\
\mathfrak{x}'\left( 1+\frac{\mathfrak{x}^{2}}{x^{2}}\right)+\frac{\mathfrak{x}\mathfrak{t}\mathfrak{t}'}{x^{2}}=0.\label{eqn45}
\end{eqnarray}


The solution corresponding to the above set of equations (\ref{eqn45}) could be formally expressed as,
\begin{eqnarray}
z(\tau)&=&\mathcal{O}\left( \frac{1}{\kappa^{2}}\right)\nonumber\\
\mathfrak{x}(\tau)&=&\mathfrak{t}(\tau)=\mathfrak{c}+\mathcal{O}\left( \frac{1}{\kappa^{2}}\right).
\end{eqnarray}

Therefore, in the non perturbative ($ \kappa \gg 1 $) regime, the leading order contribution to the on-shell action vanishes. Let us now go beyond this trivial regime and consider corrections corresponding to the next sub-leading order at large $ \kappa $. The corresponding action takes the following form, 
\begin{eqnarray}
S_{PAdS_3}\simeq\frac{\hat{T}}{2}\int  \frac{d^{2}\sigma}{x^{2}}\left( \mathfrak{x}\mathfrak{t}' -\mathfrak{t}\mathfrak{x}'\right)^{2} -\frac{\hat{T}}{2\kappa^{2}}\int \frac{d^{2}\sigma}{x^{4}} \left( x^{2}z'^{2}-z^{2}\left( \mathfrak{x}\mathfrak{t}' -\mathfrak{t}\mathfrak{x}'\right)^{2}\right)+.. 
\end{eqnarray}

Considering the following fact namely, 
\begin{eqnarray}
z'(\tau)\sim x'(\tau)\sim \mathcal{O}\left( \frac{1}{\kappa^{2}}\right)
\end{eqnarray}
it is quite trivial to notice (without solving any of the equations of motion explicitly),
\begin{eqnarray}
S_{PAdS_3}\simeq \mathcal{O}\left( \frac{1}{\kappa^{4}}\right)+..
\end{eqnarray}
which thereby possesses a vanishingly small contribution (to the full on-shell action ($ S_{AdS_3\times S^{3}} $)) as compared to that of the corresponding contribution appearing from $ S^{3} $ in the large $ \kappa (\gg 1) $ limit (as we shall see shortly). In other words, in the non perturbative regime ($ \kappa \gg 1 $), the dynamics associated with the two point correlation is largely determined by the corresponding partition function associated with the three sphere ($ S^{3} $).


\subsection{Solutions in $ S^{3} $}
We now consider the dynamics of strings in $ S^{3} $. In order to proceed further, we choose the following ansatz \cite{Park:2010vs},
\begin{eqnarray}
\theta = \theta (\varsigma),~~\varphi = \nu \tau + \mathfrak{g}(\varsigma),~~\phi =C
\end{eqnarray}
where, $ \varsigma =\mathfrak{a}\tau + \mathfrak{b}\sigma $. With this choice in hand, we essentially confine ourselves to the subspace $ S^{2} $ of the full three sphere.

In order to proceed further, we first note-down the equation of motion corresponding to $ \varphi(\varsigma) $ which yields,
\begin{eqnarray}
\partial_{\varsigma}\left(\frac{\sin^{2}\theta}{(1+\kappa^{2}\cos^{2}\theta)}(\mathfrak{a}\nu +(\mathfrak{a}^{2}-\mathfrak{b}^{2}) \mathfrak{g}'(\varsigma))\right) =0\label{E42}
\end{eqnarray}
where, the prime denotes derivative w.r.t the variable, $ \varsigma $.

Integrating the above equation (\ref{E42}) once we find,
\begin{eqnarray}
\mathfrak{g}'(\varsigma)=\frac{1}{(\mathfrak{b}^{2}-\mathfrak{a}^{2})}\left(\mathfrak{a}\nu -\frac{\mathcal{C}(1+\kappa^{2}\cos^{2}\theta)}{\sin^{2}\theta} \right) \label{e43}
\end{eqnarray}
where, $ \mathcal{C} $ is some integration constant.

We now focus our attention towards computing the equation of motion corresponding to $ \theta (\varsigma) $. In order to do that, instead of considering the dynamics directly, we turn our attention towards the first integrals of motion namely the Virasoro constraints of the theory \cite{Lee:2008ui},
\begin{eqnarray}
T_{\tau\tau}+T_{\sigma\sigma}+2T_{\tau \sigma}=0.
\end{eqnarray}

A straightforward computation yields the following,
\begin{eqnarray}
\theta'^{2}=\frac{\mathfrak{b}^{2}\nu^{2}}{(\mathfrak{b}^{2}-\mathfrak{a}^{2})^{2}\sin^{2}\theta}\left( 1+\frac{\kappa^{2}\mathcal{C}}{\mathfrak{b}\nu}\right)^{2} \left[-\sin^{4}\theta +\frac{\tilde{\mathcal{C}}^{2}(1+\kappa^{2})^{2}}{\mathfrak{b}^{2}\nu^{2}\left( 1+\frac{\kappa^{2}\mathcal{C}}{\mathfrak{b}\nu}\right)^{2}}\sin^{2}\theta -\frac{\mathcal{C}^{2}(1+\kappa^{2})^{2}}{\mathfrak{b}^{2}\nu^{2}\left( 1+\frac{\kappa^{2}\mathcal{C}}{\mathfrak{b}\nu}\right)^{2}}\right]. \label{E45}
\end{eqnarray}

As a next step of our analysis, we factorize (\ref{E45}) as,
\begin{eqnarray}
\theta'^{2}=\frac{\mathfrak{b}^{2}\nu^{2}}{(\mathfrak{b}^{2}-\mathfrak{a}^{2})^{2}\sin^{2}\theta}\left( 1+\frac{\kappa^{2}\mathcal{C}}{\mathfrak{b}\nu}\right)^{2} (\sin^{2}\theta_{max}-\sin^{2}\theta)(\sin^{2}\theta -\sin^{2}\theta_{min})
\label{e46}
\end{eqnarray}
where,
\begin{eqnarray}
\sin^{2}\theta_{max}+\sin^{2}\theta_{min}=\frac{\tilde{\mathcal{C}}^{2}(1+\kappa^{2})^{2}}{\mathfrak{b}^{2}\nu^{2}\left( 1+\frac{\kappa^{2}\mathcal{C}}{\mathfrak{b}\nu}\right)^{2}}\nonumber\\
\sin^{2}\theta_{max}\sin^{2}\theta_{min}=\frac{\mathcal{C}^{2}(1+\kappa^{2})^{2}}{\mathfrak{b}^{2}\nu^{2}\left( 1+\frac{\kappa^{2}\mathcal{C}}{\mathfrak{b}\nu}\right)^{2}}.\label{e47}
\end{eqnarray}

Notice that, here $ \theta_{max} $ and $ \theta_{min} $ correspond to extremal values of $ \theta $ such that $ \theta' =0 $. It turns out that the size of the magnon and/or spike in the dual gauge theory could be estimated by means of $ \theta_{max} $ \cite{Park:2010vs}. In our analysis, we first consider the infinite size limit associated with these single trace operators in the dual gauge theory. This infinite size limit corresponds to setting, $ \sin\theta_{max}=1 $ in the bulk. As far as the dual field theory is concerned, this infinite size limit would correspond magnons with large angular momenta together with finite angular difference (or momentum) and spikes with large angular difference between its two end points together with finite angular momentum \cite{Park:2010vs}.  

At this stage, it is noteworthy to mention that the two point correlations between these heavy states in the dual gauge theory should not follow the usual power law behavior \cite{Park:2010vs} of a CFT and in fact there should be a clear deviation from the usual power law behavior indicating the fact that the original conformal symmetry is broken. Therefore it remains as an interesting direction to be explored how does this two point correlation behave in the presence of an integrable one parameter background deformations \cite{Delduc:2013qra}. In the first part of our analysis, we precisely address this issue by analytically computing the two point correlations between two heavy magnon states. In the second part of our analysis, considering the large size limit, we compute two point correlation function between single trace operators dual to spiky constructions over the deformed background.  We also discuss the finite size corrections to these correlation functions in each of the above examples
.
\section{Magnons}
\subsection{The large size limit}
Both the infinite as well as the finite size limit for magnons correspond to setting, $ \partial_{\sigma}\varphi =0 $ at $ \theta =\theta_{max} $ \cite{Park:2010vs}. For $ \theta_{max}=\frac{\pi}{2} $, this implies a large value for the angular momenta at a finite angular difference \cite{Park:2010vs}. Using (\ref{e43}), this naturally implies,
\begin{eqnarray}
\sin^{2}\theta_{max}=\frac{\mathcal{C}(1+\kappa^{2})}{\mathfrak{a}\nu(1 +\frac{\kappa^{2}\mathcal{C}}{\mathfrak{a}\nu})}.\label{e48}
\end{eqnarray}

The infinite size limit for magnons corresponds to setting, $ \mathcal{C}=\mathfrak{a}\nu $ \cite{Park:2010vs}.
Substituting (\ref{e48}) into (\ref{e47}), this further yields,
\begin{eqnarray}
\sin^{2}\theta_{min}=\frac{\mathfrak{a}^{2}(1+\kappa^{2})^{2}}{\mathfrak{b}^{2}\left( 1+\frac{\kappa^{2}\mathfrak{a}}{\mathfrak{b}}\right)^{2}}.\label{E49}
\end{eqnarray}
Clearly, in the limit, $ \kappa \rightarrow 0 $ one recovers the results corresponding to the giant magnon solutions constructed over the un-deformed background \cite{Park:2010vs}. Using, (\ref{e48}) and (\ref{E49}) we can re write (\ref{e46}) as,
\begin{eqnarray}
\theta'^{2}=\frac{\mathfrak{b}^{2}\nu^{2}}{(\mathfrak{b}^{2}-\mathfrak{a}^{2})^{2}\sin^{2}\theta}\left( 1+\frac{\kappa^{2}\mathcal{C}}{\mathfrak{b}\nu}\right)^{2} \left( \frac{\mathcal{C}(1+\kappa^{2})}{\mathfrak{a}\nu(1 +\frac{\kappa^{2}\mathcal{C}}{\mathfrak{a}\nu})}-\sin^{2}\theta\right) \left( \sin^{2}\theta -\frac{\mathfrak{a}\mathcal{C}(1+\kappa^{2})^{2}}{\mathfrak{b}^{2}\nu\left( 1+\frac{\kappa^{2}\mathfrak{a}}{\mathfrak{b}}\right)^{2}}\right).
\label{E50}
\end{eqnarray}

The purpose of the present analysis is to perform an explicit analytic computation on two point correlations between two \textit{heavy} magnon states in the \textit{classical} limit \cite{Park:2010vs}. It turns out that in the classical limit, the path integral is dominated by the Polyakov action evaluated at the classical saddle point \cite{Janik:2010gc}. Following the original prescription \cite{Janik:2010gc}, the classical Polyakov action (corresponding to $ S^{3} $) for giant magnon solutions turns out to be,
\begin{eqnarray}
\bar{S}_{PS^{3}}=S_{PS^{3}}-\int d^{2}\sigma \left( \Pi_{\theta}\dot{\theta}+\Pi_{\varphi}\dot{\varphi}\right) 
\end{eqnarray}
where, $ \Pi _{\theta}$ and $ \Pi_{\varphi} $ are the conjugate momenta and dot corresponds to derivative w.r.t. $ \tau $. 

A straightforward computation yields,
\begin{eqnarray}
\bar{S}_{PS^{3}}=-\frac{\hat{T}}{2}\int d^{2}\sigma\left( \tilde{\mathfrak{f}}(\theta)(\mathfrak{a}^{2}+\mathfrak{b}^{2})\theta'^{2}+\tilde{\mathfrak{h}}(\theta)(\nu^{2}+2\mathfrak{a}\nu \mathfrak{g}'+(\mathfrak{a}^{2}+\mathfrak{b}^{2})\mathfrak{g}'^{2})\right). \label{E52}
\end{eqnarray}

Using (\ref{e43}) and (\ref{E50}), we finally obtain,
\begin{eqnarray}
\bar{S}_{PS^{3}}&=&-\frac{\hat{T}}{2}\int_{-s/2}^{s/2}d\tau \int_{-L}^{L}d\sigma\frac{\nu^{2}\left( 1+\frac{2\kappa^{2}\mathfrak{ab}\cos^{2}\theta}{(\mathfrak{a}+\mathfrak{b})^{2}}\right) }{(1+\kappa^{2}\cos^{2}\theta)}\nonumber\\
&=&-\hat{T}s \alpha \nu^{2}\left( 1+\frac{\kappa^{2}\mathfrak{a}}{\mathfrak{b}}\right)^{-1}\int_{\theta_{min}}^{\pi / 2} \frac{\sin\theta \left( 1+\frac{2\kappa^{2}\mathfrak{ab}\cos^{2}\theta}{(\mathfrak{a}+\mathfrak{b})^{2}}\right) }{\cos\theta(1+\kappa^{2}\cos^{2}\theta)}\frac{d\theta}{\sqrt{\sin^{2}\theta -\sin^{2}\theta_{min}}}\nonumber\\
& =& -\hat{T}s L \nu^{2}\Theta^{(M)} 
\label{E53}
\end{eqnarray}
where, the entity $ \Theta^{(M)} $ could be formally expressed as,
\begin{eqnarray}
\Theta^{(M)} = \frac{\alpha}{L}\left( 1+\frac{\kappa^{2}\mathfrak{a}}{\mathfrak{b}}\right)^{-1}\int_{\theta_{min}}^{\pi / 2} \frac{\sin\theta \left( 1+\frac{2\kappa^{2}\mathfrak{ab}\cos^{2}\theta}{(\mathfrak{a}+\mathfrak{b})^{2}}\right) }{\cos\theta(1+\kappa^{2}\cos^{2}\theta)}\frac{d\theta}{\sqrt{\sin^{2}\theta -\sin^{2}\theta_{min}}}
\end{eqnarray}
and $ \alpha $ is some numerical prefactor. Interestingly enough and unlike the $ AdS_{3} $ example, the above expression (\ref{E53}) is \textit{exact} in trms of the background deformations. Finally, it is also noteworthy to mention that, in the limit, $ \kappa \rightarrow 0 $ one could trivially convert the $ \theta $ integral into an integral over $ \sigma $ which finally yields, $ \Theta^{(M)} =1 $.

Combining (\ref{E32}) and (\ref{E53}), we finally obtain,
\begin{eqnarray}
i S_{AdS_3\times S^{3}} =  i\hat{T}sL\left(\frac{4}{s^{2}}\mathcal{Z} (s) \log^{2}|\frac{x_{f}}{\epsilon}|-\nu^{2}\Theta^{(M)} \right) \label{e55}
\end{eqnarray}
where, the entity $ \mathcal{Z}(s) $ could be formally expressed as,
\begin{eqnarray}
\mathcal{Z}(s)&=&1+\frac{\kappa^{2}}{s}\int_{-s/2}^{s/2} d\tau \mathfrak{K}(\tau)+\mathcal{O}(\kappa^{4})\nonumber\\
&=&1+\frac{\kappa^{2}}{s} \mathcal{Q}(s)+\mathcal{O}(\kappa^{4})\label{e56}
\end{eqnarray}
where, the functional form of $ \mathcal{Q}(s) $ could be uniquely fixed by means of the corresponding saddle point equation evaluated at the classical saddle point.

As a next step of our analysis, we determine the classical saddle point, $ s= \bar{s} $ which is achieved by varying the action (\ref{e55}) w.r.t. the parameter $ s $ that yields the following differential equation,
\begin{eqnarray}
-\frac{1}{s}-\frac{s\nu^{2}\Theta^{(M)} }{4 \log^{2}|\frac{x_{f}}{\epsilon}|}+\kappa^{2}\left(\frac{\mathcal{Q}(s)}{s} \right)' -\kappa^{2}\left( \frac{\mathcal{Q}(s)}{s^{2}}\right)=0.\label{e57}
\end{eqnarray}

In the following, we would work out the saddle point solutions corresponding to (\ref{e57}) upto leading order in the deformation parameter. In other words, our solutions would be valid $ \sim \mathcal{O}(\kappa^{2}) $. These solutions turn out to be,
\begin{eqnarray}
\mathcal{Q}(\bar{s})&\approx &\mathcal{Q}(\bar{s}^{(0)})=i\mathfrak{q}\bar{s}^{2(0)}+\mathcal{O}(\kappa^{2})\nonumber\\
&=&-\frac{4i \mathfrak{q}}{\nu^{2}}\log^{2} |\frac{x_{f}}{\epsilon}|+\mathcal{O}(\kappa^{2})\nonumber\\
&=&-\frac{i\mathfrak{q}\beta^{2}\bar{s}^{2(0)}}{\nu^{2}}+\mathcal{O}(\kappa^{2}) \nonumber\\
\bar{s}&\approx &-\frac{2i}{\nu \sqrt{\Theta^{(M)}}}\log |\frac{x_{f}}{\epsilon}|\label{e58}
\end{eqnarray}
where, $ \mathfrak{q} $ is the integration constant.

Substituting (\ref{e58}) into (\ref{e55}), the semi-classical partition function evaluated at the classical saddle point turns out to be,
\begin{eqnarray}
i S_{AdS_3\times S^{3}}\approx \left(1+\frac{\kappa^{2} \mathfrak{q}}{\nu } \log |\frac{x_{f}}{\epsilon}|\right)\log | \frac{\epsilon}{x_{f}}|^{2 \Delta_{\kappa}} \label{e59}
\end{eqnarray}
which in the limit, $ \kappa \rightarrow 0$ precisely matches to that with the standard formula corresponding to the semi classical string partition function estimated over the $ AdS_5 \times  S^{2}$ background \cite{Park:2010vs}. Here, the entity,
\begin{eqnarray}
\Delta_{\kappa} =\hat{T} \int_{-L}^{L} d\sigma \nu\sqrt{\Theta^{(M)}}
\end{eqnarray}
could be thought of as being that of the reminiscent of the classical conformal dimension associated with heavy single trace operators in the dual gauge theory. In the limit, $ \kappa \rightarrow 0 $ this precisely matches with the classical conformal dimension associated with single trace operators dual to long stringy solutions in the bulk.

As a consistency check of our analysis, below we show that in the appropriate limit ($ \kappa \rightarrow 0$), our result trivially reduces to that of the giant magnon dispersion relation at strong coupling \cite{Hofman:2006xt} where one could easily identify the entity $ \Delta_{\kappa =0} $ as being that of the classical energy associated with stringy excitation in the bulk \cite{Park:2010vs}. 

We first compute the angular momentum,
\begin{eqnarray}
J_{\varphi}=2\hat{T}\int_{\theta_{min}}^{\pi /2}\frac{\sin \theta}{\cos \theta \sqrt{\sin^{2}\theta -\sin^{2}\theta_{min}}}\frac{\left( 1+\frac{\kappa^{2}\mathfrak{a}}{\mathfrak{b}}\right)^{-1} }{(1+\kappa^{2}\cos^{2}\theta)}\left(\sin^{2}\theta -\frac{\mathfrak{a}^{2}}{\mathfrak{b}^{2}}(1+\kappa^{2}\cos^{2}\theta) \right)d\theta \end{eqnarray}
which clearly diverges in the limit, $ \theta \rightarrow \pi /2$. However, the difference 
\begin{eqnarray}
\Delta_{\kappa}-J_{\varphi}=2 \hat{T}\int_{\theta_{min}}^{\pi /2}\frac{\sin \theta}{\cos \theta \sqrt{\sin^{2}\theta -\sin^{2}\theta_{min}}}\mathfrak{F}(\kappa^{2},\theta)d\theta \label{E59}
\end{eqnarray}
where, the function $\mathfrak{F}(\kappa^{2},\theta)$ could be formally expressed as,
\begin{eqnarray}
\mathfrak{F}(\kappa^{2},\theta)=\sqrt{\Theta^{(M)}}\left( 1+\frac{2\kappa^{2}\mathfrak{a}}{(\mathfrak{a}+\mathfrak{b})}\right)^{-1}\left( 1+\frac{\kappa^{2}\mathfrak{a}}{\mathfrak{b}}\right)\cos^{2}\theta_{min}\nonumber\\-\frac{\left( 1+\frac{\kappa^{2}\mathfrak{a}}{\mathfrak{b}}\right)^{-1} }{(1+\kappa^{2}\cos^{2}\theta)}\left(\sin^{2}\theta -\frac{\mathfrak{a}^{2}}{\mathfrak{b}^{2}}(1+\kappa^{2}\cos^{2}\theta) \right).
\end{eqnarray}
It turns out that in the limit, $ \kappa \rightarrow 0 $
\begin{eqnarray}
\Delta_{\kappa =0}-J_{\varphi}=2 T\int_{\theta_{min}}^{\pi /2}\frac{\sin \theta \cos \theta}{ \sqrt{\sin^{2}\theta -\sin^{2}\theta_{min}}}d\theta = 2T \cos\theta_{min} \label{E61}
\end{eqnarray}
which is finite as expected \cite{Hofman:2006xt}.

On the other hand, the angle difference between the two end points of the string turns out to be,
\begin{eqnarray}
\Delta \varphi \equiv p =2\int_{\theta_{min}}^{\pi /2}\frac{\mathfrak{a}}{\mathfrak{b}}\left(1+\frac{\kappa^{2}\mathfrak{a}}{\mathfrak{b}} \right)^{-1} \frac{(1+\kappa^{2})\cos\theta}{\sin\theta \sqrt{\sin^{2}\theta -\sin^{2}\theta_{min}}}d\theta \nonumber\\
=\frac{2}{\sin\theta_{min}}\frac{\mathfrak{a}}{\mathfrak{b}}\left(1+\frac{\kappa^{2}\mathfrak{a}}{\mathfrak{b}} \right)^{-1}(1+\kappa^{2})\sin^{-1}(\cos\theta_{min})
\end{eqnarray}
which trivially reduces to,
\begin{eqnarray}
\Delta \varphi \equiv p 
=2\sin^{-1}(\cos\theta_{min})
\end{eqnarray}
in the limit, $ \kappa  \rightarrow 0$. In summary, the dispersion relation \cite{Hofman:2006xt},\cite{Park:2010vs},
\begin{eqnarray}
\Delta_{\kappa =0}-J_{\varphi}=2 T |\sin \frac{p}{2}|\label{Eqn62}
\end{eqnarray}
is trivially satisfied in the limit, $ \kappa \rightarrow 0 $. However, the above relation (\ref{Eqn62}) does not hold in the presence of background deformations and it receives non trivial corrections namely,
\begin{eqnarray}
\Delta_{\kappa}-J_{\varphi}=2 \hat{T}\int_{\theta_{min}}^{\pi /2}\frac{\sin \theta}{\cos \theta \sqrt{\sin^{2}\theta -1+\sin^{2}\Delta \tilde{\varphi}}}\mathfrak{F}(\kappa^{2},\theta)d\theta
\end{eqnarray}
where,
\begin{eqnarray}
\mathfrak{F}(\kappa^{2},\theta)=\sqrt{\Theta^{(M)}}\left( 1+\frac{2\kappa^{2}\mathfrak{a}}{(\mathfrak{a}+\mathfrak{b})}\right)^{-1}\left( 1+\frac{\kappa^{2}\mathfrak{a}}{\mathfrak{b}}\right)\sin^{2}\Delta \tilde{\varphi}\nonumber\\-\frac{\left( 1+\frac{\kappa^{2}\mathfrak{a}}{\mathfrak{b}}\right)^{-1} }{(1+\kappa^{2}\cos^{2}\theta)}\left(\sin^{2}\theta -\frac{\mathfrak{a}^{2}}{\mathfrak{b}^{2}}(1+\kappa^{2}\cos^{2}\theta) \right)
\end{eqnarray}
together with the fact that,
\begin{eqnarray}
\Delta \tilde{\varphi}=\left( \frac{2\mathfrak{a}}{\mathfrak{b}\sin\theta_{min}}\right) ^{-1}\left(1+\frac{\kappa^{2}\mathfrak{a}}{\mathfrak{b}} \right)(1+\kappa^{2})^{-1}\Delta\varphi.
\end{eqnarray}

Therefore, the excitation associated with the gauge theory dual to $ \kappa $- deformed background are not magnons in the usual sense. As a natural consequence of this, it is also not quite confirmed whether the corresponding spin chain description holds for the dual gauge theory in the limit of weak couplings.  

We now compute two point function between magnon like excitations in the dual gauge theory. Following the original prescription \cite{Janik:2010gc}, the corresponding two point correlation between the heavy magnon like states finally turns out to be,
\begin{eqnarray}
\langle \mathcal{O}(0)\mathcal{O}(x_{f})\rangle_{\kappa}  = e^{i S_{AdS_3\times S^{3}}}\approx \left( \frac{\epsilon}{x_{f}}\right)^{2\Delta_{\kappa}}~  e^{-\frac{2\kappa^{2}\mathfrak{q}\Delta_{\kappa}}{\nu}\log^{2}|\frac{\epsilon}{x_{f}} |+\mathcal{O}(\kappa^{4})}\label{e61}
\end{eqnarray}
where, the exponential suppression above in the two point function (\ref{e61}) has its origin in the point particle dynamics within the deformed $ AdS_{3} $ sector. On the other hand, the associated power law behavior is exact in the deformation and its origin lies entirely on the dynamics of strings over the deformed sphere.  It is therefore quite tempting to claim that (\ref{e61}) is a \textit{semi-perturbative} result in itself.

The above relation (\ref{e61}) is also quite intuitive in the sense that for a given separation ($\delta x= x_{f} $), the leading contribution to the two point correlations between heavy magnon like states is exponentially suppressed compared to that of their cousins in the gauge theory ($ CFT_2 $) dual to undeofrmed $ AdS_3 \times S^{3}$. In other words, in the presence of background deformations, the leading contribution to the two point correlation function between two heavy operators dies at rate faster compared to that of the original unreformed theory. 

Before we finish our discussion, it is worth specifying the corresponding behavior of the two point correlation in the non perturbative regime, namely for $ \kappa \gg 1$. It turns out that in the non perturbative ($ \kappa \gg 1 $) regime,
\begin{eqnarray}
\bar{S}_{PS^{3}}\simeq \frac{i\hat{T}s \mathfrak{C}^{(M)} }{\kappa^{2}}+\mathcal{O}\left( \frac{1}{\kappa^{4}}\right)\label{eqn81}
\end{eqnarray}
where, $ \mathfrak{C}^{(M)}=\frac{2 \varepsilon \nu (\mathfrak{b}-\mathfrak{a}) }{(\mathfrak{b}+\mathfrak{a}) }\frac{\sin \varepsilon}{\cos^{2}\varepsilon} $ and, $ |\varepsilon|\ll 1 $. 

The above equation (\ref{eqn81}) clearly exhibits the fact that in the non perturbative regime ($ \kappa \gg 1 $), the dominant contribution to the two point correlation appears from the sphere partition function. This might be regarded as the consequence of the fact that in the limit, $ \kappa \gg 1 $, the allowed physical region for strings moving in $ AdS_{3} $ eventually shrinks to zero \cite{Hoare:2014pna}. 

Using (\ref{eqn81}), the two point correlation function associated with giant magnon like states in the dual gauge theory could be formally expressed as,
\begin{eqnarray}
\langle \mathcal{O}(0)\mathcal{O}(x_{f})\rangle_{\kappa \rightarrow \infty}\approx e^{-\frac{\hat{T}\bar{s} \mathfrak{C}^{(M)} }{\kappa^{2}}+\mathcal{O}\left( \frac{1}{\kappa^{4}}\right)}.
\end{eqnarray}

\subsection{Finite size corrections}
Having done our explorations on giant magnon two point correlations, the purpose of this Section is to pursue the same computation in the finite charge limit. In other words, we compute two point correlation function between single trace operators of finite length. In order to do that, the first thing we consider is to set $ \theta_{max}\neq \frac{\pi}{2} $ which in turn implies that, 
\begin{eqnarray}
\sin^{2}\theta_{max}&=&\frac{\mathcal{C}(1+\kappa^{2})}{\mathfrak{a}\nu(1 +\frac{\kappa^{2}\mathcal{C}}{\mathfrak{a}\nu})}\nonumber\\
\sin^{2}\theta_{min}&=&\frac{\mathfrak{a}\mathcal{C}(1+\kappa^{2})}{\mathfrak{b}^{2}\nu}\frac{\left(1+\frac{\kappa^{2}\mathcal{C}}{\mathfrak{a}\nu}\right)}{\left(1+\frac{\kappa^{2}\mathcal{C}}{\mathfrak{b}\nu}\right)^{2}}\label{E63}
\end{eqnarray}
such that, $ \mathcal{C}\neq \mathfrak{a}\nu $. As a consequence of this we note,
\begin{eqnarray}
\theta'^{2}=\frac{\mathfrak{b}^{2}\nu^{2}}{(\mathfrak{b}^{2}-\mathfrak{a}^{2})^{2}\sin^{2}\theta}\left( 1+\frac{\kappa^{2}\mathcal{C}}{\mathfrak{b}\nu}\right)^{2} \left( \frac{\mathcal{C}(1+\kappa^{2})}{\mathfrak{a}\nu(1 +\frac{\kappa^{2}\mathcal{C}}{\mathfrak{a}\nu})}-\sin^{2}\theta\right) \left( \sin^{2}\theta -\frac{\mathfrak{a}\mathcal{C}(1+\kappa^{2})}{\mathfrak{b}^{2}\nu}\frac{\left(1+\frac{\kappa^{2}\mathcal{C}}{\mathfrak{a}\nu}\right)}{\left(1+\frac{\kappa^{2}\mathcal{C}}{\mathfrak{b}\nu}\right)^{2}}\right).
\label{E66}
\end{eqnarray}

Our next task would be to compute the on-shell action (\ref{E52}) corresponding to deformed $ S^{3} $ which for the present case yields,
\begin{eqnarray}
\bar{S}_{PS^{3}}&=&-\frac{\hat{T}}{2}\int_{-s/2}^{s/2}d\tau \int_{-L}^{L}d\sigma ~ \frac{\mathcal{C}\nu}{\mathfrak{a}}\frac{\varrho (\theta)}{(1+\kappa^{2}\cos^{2}\theta)} \nonumber\\
&=&-\hat{T}sL\frac{\mathcal{C}\nu}{\mathfrak{a b}}\Theta^{(M)}_{F}=-\hat{T}sL \nu^{2}_{F}\Theta^{(M)}_{F}\nonumber\\
\Theta^{(M)}_{F}&=&\frac{1}{L}\int_{\theta_{min}}^{\theta_{max}} \frac{\varrho(\theta)}{\theta '(1+\kappa^{2}\cos^{2}\theta)} d\theta \label{E64}
\end{eqnarray}
where, the entity $ \varrho (\theta) $ could be formally expressed as,
\begin{eqnarray}
\varrho (\theta)=\frac{(\mathfrak{a}^{2}+\mathfrak{b}^{2})(1+\kappa^{2})}{(\mathfrak{b}^{2}-\mathfrak{a}^{2})^{2}\left( 1+\frac{\kappa^{2}\mathcal{C}}{\mathfrak{a}\nu}\right) }\left[ \mathfrak{b}^{2}\left( 1+\frac{\kappa^{2}\mathcal{C}}{\mathfrak{b}\nu}\right)^{2}+\mathfrak{a}^{2}\left( 1+\frac{\kappa^{2}\mathcal{C}}{\mathfrak{a}\nu}\right)^{2}\right]-\frac{4 \mathfrak{a}^{2}\mathfrak{b}^{2}(1+\kappa^{2}\cos^{2}\theta)}{(\mathfrak{b}^{2}-\mathfrak{a}^{2})^{2}}\nonumber\\
-\frac{2\kappa^{2}\mathfrak{a}\mathfrak{b}(\mathfrak{a}^{2}+\mathfrak{b}^{2})}{(\mathfrak{b}^{2}-\mathfrak{a}^{2})^{2}} \left[\sin^{2}\theta+\frac{\mathcal{C}}{\mathfrak{b}\nu}(1+\kappa^{2}) \right]. 
\end{eqnarray}

Two points are to be noted here. First of all, the above expression (\ref{E64}) is an exact result in terms of the background deformation and secondly it produces the correct result \cite{Lee:2011fe} in the limit, $ \kappa \rightarrow 0 $.

As a next step, we write down the total action,
\begin{eqnarray}
i S_{AdS_{3}\times S^{3}}=i\hat{T}sL\left(\frac{4}{s^{2}}\mathcal{Z} (s) \log^{2}|\frac{x_{f}}{\epsilon}|-\nu_{F}^{2}\Theta^{(M)}_{F} \right).\label{E66}
\end{eqnarray}

Following the same steps as earlier and evaluating (\ref{E66}) at the classical saddle point we find,
\begin{eqnarray}
i S_{AdS_3\times S^{3}}\approx \left(1+\frac{\kappa^{2} \mathfrak{q}}{\nu_{F} } \log |\frac{x_{f}}{\epsilon}|\right)\log | \frac{\epsilon}{x_{f}}|^{2 \Delta^{(F)}_{\kappa}}  \label{e67}
\end{eqnarray}
where, the entity,
\begin{eqnarray}
\Delta^{(F)}_{\kappa} =\hat{T} \int_{-L}^{L} d\sigma \nu_{F}\sqrt{\Theta^{(M)}_{F}}
\end{eqnarray} 
becomes exactly the classical conformal dimension associated with finite sized single trace (magnons) operators in the limit, $ \kappa \rightarrow 0 $ \cite{Lee:2011fe}.

Finally, we note down the corresponding two point correlation function between single trace operators of finite size which takes the following form,
\begin{eqnarray}
\langle \mathcal{O}(0)\mathcal{O}(x_{f})\rangle_{\kappa}  \approx \left( \frac{\epsilon}{x_{f}}\right)^{2\Delta^{(F)}_{\kappa}}~  e^{-\frac{2\kappa^{2}\mathfrak{q}\Delta^{(F)}_{\kappa}}{\nu_{F}}\log^{2}|\frac{\epsilon}{x_{f}} |+\mathcal{O}(\kappa^{4})}.\label{E69}
\end{eqnarray}
Therefore, compared with the previous example, the only difference that we encounter here is that the entities like, $ \Delta_{\kappa} $ and $ \nu $ are now corrected due to the finite size effects.

Like in the previous section, we now focus on the large $ \kappa(\gg 1) $ behavior of two point function in the limit of finite size corrections where we note,
\begin{eqnarray}
\bar{S}_{PS^{3}}\simeq \frac{i\hat{T}s \mathfrak{C}^{(M)}_{F} }{\kappa^{2}}+\mathcal{O}\left( \frac{1}{\kappa^{4}}\right)\label{eqn92}
\end{eqnarray}
where, $ \mathfrak{C}^{(M)}_{F}= \frac{4\varepsilon\nu^{2}_{F}\mathfrak{a}^{2}\mathfrak{b}^{2}}{(\mathfrak{b}^{2} -\mathfrak{a}^{2})\mathcal{C}}\frac{\sin\varepsilon}{\cos^{2}\varepsilon} $.

Using, (\ref{eqn92}), the finite size corrections to two point correlation function in the limit of large background deformations could be formally expressed as,
\begin{eqnarray}
\langle \mathcal{O}(0)\mathcal{O}(x_{f})\rangle_{\kappa \rightarrow \infty}\approx e^{-\frac{\hat{T}\bar{s} \mathfrak{C}^{(M)}_{F} }{\kappa^{2}}+\mathcal{O}\left( \frac{1}{\kappa^{4}}\right)}.
\end{eqnarray}
\section{Spikes}
\subsection{The large size limit}
Having done our computations on two point correlations between single trace (magnon) operators, we now turn our attention towards the computation of two point correlation function between operators dual to spiky string solutions over the $ \kappa $-deformed background. We first consider the large size limit which corresponds to a large angular difference between the two end points of the spike \cite{Park:2010vs}. The corresponding boundary condition that one uses is, $ \partial_{\tau}\varphi =0 $ at $ \theta =\theta_{max} $ \cite{Park:2010vs} which for the present case yields,
\begin{eqnarray}
\sin^{2}\theta_{max}&=&\frac{\mathfrak{a}\mathcal{C}(1+\kappa^{2})}{\mathfrak{b}^{2}\nu \left(1+\frac{\kappa^{2}\mathfrak{a}\mathcal{C}}{\mathfrak{b}^{2}\nu} \right) }\nonumber\\
\sin^{2}\theta_{min}&=&\frac{\mathcal{C}(1+\kappa^{2})\left(1+\frac{\kappa^{2}\mathfrak{a}\mathcal{C}}{\mathfrak{b}^{2}\nu} \right) }{\mathfrak{a}\nu \left( 1+\frac{\kappa^{2}\mathcal{C}}{\mathfrak{b}\nu}\right)^{2} }.
\end{eqnarray}
The large size limit corresponds to setting, $ \sin^{2}\theta_{max}=1 $ which in turn implies, $ \mathcal{C}=\frac{\mathfrak{b}^{2}\nu}{\mathfrak{a}} $. 

Using this value for the constant we find,
\begin{eqnarray}
\sin^{2}\theta_{min}=\frac{\mathfrak{b}^{2}(1+\kappa^{2})^{2}}{\mathfrak{a}^{2}\left( 1+\frac{\kappa^{2}\mathfrak{b}}{\mathfrak{a}}\right)^{2}}.
\end{eqnarray}

Finally, the on-shell action (\ref{E52}) corresponding to deformed $ S^{3} $ turns out to be,
\begin{eqnarray}
\bar{S}_{PS^{3}}&=&-\frac{\hat{T}}{2}\int_{-s/2}^{s/2}d\tau \int_{-L}^{L}d\sigma ~ \frac{\mathfrak{b}^{2}\nu^{2}}{(\mathfrak{b}^{2}-\mathfrak{a}^{2})^{2}}\frac{\xi (\theta)}{(1+\kappa^{2}\cos^{2}\theta)} \nonumber\\
&=&-\hat{T}sL\frac{\mathfrak{b}\nu^{2}}{(\mathfrak{b}^{2}-\mathfrak{a}^{2})^{2}}\Theta^{(S)}=-\hat{T}sL \tilde{\nu}^{2}\Theta^{(S)}\nonumber\\
\Theta^{(S)}&=&\frac{1}{L}\int_{\theta_{min}}^{\pi / 2} \frac{\xi(\theta)}{\theta '(1+\kappa^{2}\cos^{2}\theta)} d\theta \label{E72}
\end{eqnarray}
where, the entity $ \xi(\theta) $ could be formally expressed as,
\begin{eqnarray}
\xi (\theta)=(\mathfrak{a}^{2}+\mathfrak{b}^{2})\left[ \left(1+\frac{\kappa^{2}\mathfrak{b}}{\mathfrak{a}} \right)^{2}\cos^{2}\theta -\frac{\mathfrak{b}^{2}}{\mathfrak{a}^{2}}\frac{\cos^{2}\theta}{\sin^{2}\theta}(1+\kappa^{2})^{2}\right] +\frac{\mathfrak{b}^{2}\cos^{4}\theta}{\sin^{2}\theta}(1+\kappa^{2})^{2}\nonumber\\
+\mathfrak{a}^{2}\sin^{2}\theta -2\mathfrak{b}^{2}(1+\kappa^{2}\cos^{2}\theta)+\frac{\mathfrak{b}^{4}}{\mathfrak{a}^{2}}\frac{(1+\kappa^{2}\cos^{2}\theta)^{2}}{\sin^{2}\theta}.
\end{eqnarray}

It is now quite trivial to check that in the limit, $ \kappa \rightarrow 0 $ one finds,
\begin{eqnarray}
\xi (\theta)=\frac{(\mathfrak{b}^{2}-\mathfrak{a}^{2})^{2}}{\mathfrak{a}^{2}}
\end{eqnarray}
which upon substitution into (\ref{E72}), clearly reproduces the previous results of \cite{Park:2010vs}.

Following almost similar steps as that for the magnons, the full Polyakov action turns out to be,
\begin{eqnarray}
i S_{AdS_3\times S^{3}} =  i\hat{T}sL\left(\frac{4}{s^{2}}\mathcal{Z} (s) \log^{2}|\frac{x_{f}}{\epsilon}|-\tilde{\nu}^{2}\Theta^{(S)} \right). \label{E75}
\end{eqnarray}

Evaluating (\ref{E75}) at the classical saddle point, the two point correlation function between single trace operators corresponding to the dual field theory turns out to be,
\begin{eqnarray}
\langle \mathcal{O}(0)\mathcal{O}(x_{f})\rangle_{\kappa}  \approx \left( \frac{\epsilon}{x_{f}}\right)^{2\Delta_{\kappa}}~  e^{-\frac{2\kappa^{2}\mathfrak{q}\Delta_{\kappa}}{\tilde{\nu}}\log^{2}|\frac{\epsilon}{x_{f}} |+\mathcal{O}(\kappa^{4})}\label{E61}
\end{eqnarray}
where, the entity $ \Delta_{\kappa} $ associated with spikes turns out to be,
\begin{eqnarray}
\Delta_{\kappa} =\hat{T} \int_{-L}^{L} d\sigma \tilde{\nu}\sqrt{\Theta^{(S)}}.
\end{eqnarray}
As expected, the qualitative behavior of the two point correlation (\ref{E61}) does not change compared to that with the previous example with magnons. On the other hand, the only difference between magnons and spikes appears to be in the coefficient, $ \Delta_{\kappa} $ which in the limit, $ \kappa \rightarrow 0 $ precisely matches to that with the classical conformal dimension associated with the single trace operator in the dual gauge theory.

Like in the case for magnons, we now compute two point function in the limit of large background deformations. A straightforward computation yields the following,
\begin{eqnarray}
\bar{S}_{PS^{3}}\simeq \frac{i\hat{T}s \mathfrak{C}^{(S)} }{\kappa^{2}}+\mathcal{O}\left( \frac{1}{\kappa^{4}}\right)\label{eqn102}
\end{eqnarray}
where, $ \mathfrak{C}^{(S)}= \frac{2\varepsilon\tilde{\nu}^{2}\mathfrak{a}(\mathfrak{b}^{2} -\mathfrak{a}^{2})}{\nu}\frac{\sin\varepsilon}{\cos^{2}\varepsilon} $.

Using (\ref{eqn102}), it is now indeed trivial to compute the corresponding two point function which turns out to be,
\begin{eqnarray}
\langle \mathcal{O}(0)\mathcal{O}(x_{f})\rangle_{\kappa \rightarrow \infty}\approx e^{-\frac{\hat{T}\bar{s} \mathfrak{C}^{(S)}}{\kappa^{2}}+\mathcal{O}\left( \frac{1}{\kappa^{4}}\right)}.
\end{eqnarray}
\subsection{Finite size corrections}
Like in the case for magnons, we consider, $ \theta_{max}\neq \frac{\pi}{2} $ in order to explore the effects associated with the finite size corrections on two point correlation function between single trace operators in the dual gauge theory. 

With this choice in hand, we note,
\begin{eqnarray}
\sin^{2}\theta_{max}&=&\frac{\mathfrak{a}\mathcal{C}(1+\kappa^{2})}{\mathfrak{b}^{2}\nu \left(1+\frac{\kappa^{2}\mathfrak{a}\mathcal{C}}{\mathfrak{b}^{2}\nu} \right) }\nonumber\\
\sin^{2}\theta_{min}&=&\frac{\mathcal{C}(1+\kappa^{2})\left(1+\frac{\kappa^{2}\mathfrak{a}\mathcal{C}}{\mathfrak{b}^{2}\nu} \right) }{\mathfrak{a}\nu \left( 1+\frac{\kappa^{2}\mathcal{C}}{\mathfrak{b}\nu}\right)^{2} }
\end{eqnarray}
together with the expression for,
\begin{eqnarray}
\theta'^{2}=\frac{\mathfrak{b}^{2}\nu^{2}}{(\mathfrak{b}^{2}-\mathfrak{a}^{2})^{2}\sin^{2}\theta}\left( 1+\frac{\kappa^{2}\mathcal{C}}{\mathfrak{b}\nu}\right)^{2} \left( \frac{\mathfrak{a}\mathcal{C}(1+\kappa^{2})}{\mathfrak{b}^{2}\nu \left(1+\frac{\kappa^{2}\mathfrak{a}\mathcal{C}}{\mathfrak{b}^{2}\nu} \right) }-\sin^{2}\theta\right)\nonumber\\ \left( \sin^{2}\theta -\frac{\mathcal{C}(1+\kappa^{2})\left(1+\frac{\kappa^{2}\mathfrak{a}\mathcal{C}}{\mathfrak{b}^{2}\nu} \right) }{\mathfrak{a}\nu \left( 1+\frac{\kappa^{2}\mathcal{C}}{\mathfrak{b}\nu}\right)^{2} }\right)
\label{E79}
\end{eqnarray}
such that the coefficient, $ \mathcal{C}\neq \frac{\mathfrak{b}^{2}\nu}{\mathfrak{a}} $.

Using (\ref{E79}), the on-shell action (\ref{E52}) corresponding to deformed $ S^{3} $ turns out to be,
\begin{eqnarray}
\bar{S}_{PS^{3}}&=&-\frac{\hat{T}}{2}\int_{-s/2}^{s/2}d\tau \int_{-L}^{L}d\sigma ~ \frac{\mathfrak{b}^{2}\nu^{2}}{(\mathfrak{b}^{2}-\mathfrak{a}^{2})^{2}}\frac{\xi_{F} (\theta)}{(1+\kappa^{2}\cos^{2}\theta)} \nonumber\\
&=&-\hat{T}sL\frac{\mathfrak{b}\nu^{2}}{(\mathfrak{b}^{2}-\mathfrak{a}^{2})^{2}}\Theta_{F}^{(S)}=-\hat{T}sL \tilde{\nu}^{2}\Theta_{F}^{(S)}\nonumber\\
\Theta_{F}^{(S)}&=&\frac{1}{L}\int_{\theta_{min}}^{\theta_{max}} \frac{\xi_{F}(\theta)}{\theta '(1+\kappa^{2}\cos^{2}\theta)} d\theta \label{E80}
\end{eqnarray}
where, the entity $ \xi_{F}(\theta) $ could be formally expressed as,
\begin{eqnarray}
\xi_{F} (\theta)=\frac{(\mathfrak{a}^{2}+\mathfrak{b}^{2})}{\sin^{2}\theta}\left( 1+\frac{\kappa^{2}\mathcal{C}}{\mathfrak{b}\nu}\right)^{2} \left( \frac{\mathfrak{a}\mathcal{C}(1+\kappa^{2})}{\mathfrak{b}^{2}\nu \left(1+\frac{\kappa^{2}\mathfrak{a}\mathcal{C}}{\mathfrak{b}^{2}\nu} \right) }-\sin^{2}\theta\right) \left( \sin^{2}\theta -\frac{\mathcal{C}(1+\kappa^{2})\left(1+\frac{\kappa^{2}\mathfrak{a}\mathcal{C}}{\mathfrak{b}^{2}\nu} \right) }{\mathfrak{a}\nu \left( 1+\frac{\kappa^{2}\mathcal{C}}{\mathfrak{b}\nu}\right)^{2} }\right)\nonumber\\+\frac{(\mathfrak{a}^{2}-\mathfrak{b}^{2})^{2}}{\mathfrak{b}^{2}}\sin^{2}\theta
+\frac{(\mathfrak{a}^{2}+\mathfrak{b}^{2})\sin^{2}\theta}{\mathfrak{b}^{2}\nu^{2}} \left(\mathfrak{a}\nu-\frac{\mathcal{C}(1+\kappa^{2}\cos^{2}\theta)}{ \sin^{2}\theta}\right)^{2} +\frac{2\mathfrak{a}\sin^{2}\theta}{\mathfrak{b}^{2}\nu}\left(\mathfrak{a}\nu -\frac{\mathcal{C}(1+\kappa^{2}\cos^{2}\theta)}{\sin^{2}\theta} \right).
\end{eqnarray}

Finally, following the same steps as in the previous examples, the two point correlation function between single trace operators takes the following form,
\begin{eqnarray}
\langle \mathcal{O}(0)\mathcal{O}(x_{f})\rangle_{\kappa}  \approx \left( \frac{\epsilon}{x_{f}}\right)^{2\Delta_{\kappa}}~  e^{-\frac{2\kappa^{2}\mathfrak{q}\Delta_{\kappa}}{\tilde{\nu}}\log^{2}|\frac{\epsilon}{x_{f}} |+\mathcal{O}(\kappa^{4})}\label{E82}
\end{eqnarray}
where, the entity $ \Delta_{\kappa} $ corresponding to operators dual to spiky solutions turns out to be,
\begin{eqnarray}
\Delta_{\kappa} =\hat{T} \int_{-L}^{L} d\sigma \tilde{\nu}\sqrt{\Theta_{F}^{(S)}}.
\end{eqnarray}

To conclude our discussion on spikes, following our previous methodology, we compute two point function in the limit of large background deformations ($ \kappa \gg 1 $) which for the present case yields,
\begin{eqnarray}
\langle \mathcal{O}(0)\mathcal{O}(x_{f})\rangle_{\kappa \rightarrow \infty}\approx e^{-\frac{\hat{T}\bar{s} \mathfrak{C}^{(S)}_{F}}{\kappa^{2}}+\mathcal{O}\left( \frac{1}{\kappa^{4}}\right)}
\end{eqnarray}
where, $ \mathfrak{C}^{(S)}_{F}= \frac{2\varepsilon\tilde{\nu}^{2}\mathfrak{a}(\mathfrak{b}^{2} -\mathfrak{a}^{2})}{\mathfrak{b}^{2}\nu}\frac{\sin\varepsilon}{\cos^{2}\varepsilon} $.

\section{Summary and final remarks}
We conclude our paper by mentioning some its possible future extensions that one might wish to explore. Before going into that, we first summarize the entire analysis performed so far. The goal of the present paper was to explore the underlying symmetries associated with the \textit{mysterious} dual gauge theory description corresponding to $ \kappa $- deformed $ AdS_{3}\times S^{3} $ background at strong coupling. We address this issue through a systematic computation of two point correlations between local operators at strong coupling. Our analysis is based on the basic principle \cite{Janik:2010gc}, that relates every local operator in a gauge theory to that with the (semi)classical sting states propagating over the curved geometry. 

In order to compute the two point function at strong coupling, we first solve the corresponding stringy dynamics within the physical region bounded by the so called holographic screen \cite{Kameyama:2014vma}-\cite{Kameyama:2014via} in deformed $ AdS_3 $. In our analysis, we consider two classes of local operators in the dual gauge theory namely, the magnons and the spikes. It turns out that one could solve the dynamics for strings quite exactly in deformed $ S^{3} $. However, it does not quite work that well for the deformed $ AdS_{3} $ sector. Considering both of these scenarios together, we finally able to probe the behavior of the two point function corresponding to two extremal limits of background deformations namely, the (semi)perturbative as well as the non perturbative ($ \kappa \gg 1 $) regime. Our results could therefore be extrapolated further towards the interpolating region in order to have a full qualitative understanding on the behavior of the two point function for generic background deformations. 

Our analysis reveals that in the limit of small background deformations ($ 0<\kappa \leq 1 $) associated with the deformed $ AdS_3 $ sector of the full background geometry, the corresponding two point correlation function between single trace operators in the dual gauge theory is \textit{exponentially} suppressed and as a result it decays at a rate faster than that what is expected in the usual framework of a CFT \cite{Park:2010vs}. This indeed confirms that the associated conformal invariance in the dual gauge theory is explicitly broken and also clarifies all the previous arguments \cite{Hoare:2014pna} in favor of this observation. One should further take a note on the fact that for strings attached to the holographic screen the correlation function eventually vanishes due to large exponential suppresion.

Considering the other limit namely, $ \kappa \gg 1 $ we observe that at \textit{leading} order in the background deformations, the corresponding two point function receives contributions from $ S^{3} $ and the contribution from $ AdS_{3} $ appears only at \textit{subleading} order and which is thereby suppressed compared to that of the sphere contribution. This eventually results in the saturation of the corresponding two point function at large background deformations.

Before we conclude finally, it is worth emphasizing that answers to various doubts and/or confusions associated with $ \eta $ - deformed models are not yet upto the mark. It still remains as a matter of debate whether the deformed sigma model leads towards any type IIB string theory after all. In this paper, instead of addressing this issue, we choose a rather different question to be addressed namely whether the usual notion of Gauge/String duality makes sense corresponding to classical target space solutions associated with $ \eta $- deformed model. As far as two point correlation (between heavy operators) is concerned, we find sensible answers. However, many questions are yet to be addressed that one might wish to explore in the future: (1) It would be really nice to uplift the present calculation for $ \kappa $- deformed $ AdS_5 \times S^{5} $ superstring model in the presence of non vanishing background $ B $ field,  (2) A systematic computation of the three point correlations between local operators might shed further light on the symmetries associated with the mysterious dual gauge theory description at strong coupling. (3) The present analysis could also be performed in the presence of Lax pairs. It would also be nice to compute two point functions for backgrounds without integrable deformations for example black hole geometries where the dual field theory exhibits some suitable IR cutoff. We leave these issues for future investigations.\\ \\
{\bf {Acknowledgements :}}
This work was supported by the Israeli Science Foundation with Grant Number 1635/16. The author would also like to acknowledge the financial support from the Kreitman School Of Advanced Graduate Studies, The Ben-Gurion University (BGU) of The Negev, Israel.\\ \\
{ \large{ {\bf {Appendix: Solutions for $ \mathcal{G}_{(z,x)}(\tau,\tau') $}}}}\\
 
 Here, in the Appendix, we provide the classical analysis regarding the most general solution corresponding to the Green's functions (\ref{E31}) associated with the point particle dynamics in deformed $ AdS_{3} $. We consider the most general equation of the following form,
 \begin{eqnarray}
 (\mathcal{D}_{\tau}+\mathfrak{k}^{2}(\tau))\mathcal{G}(\tau ,\tau')=-\delta (\tau -\tau')\label{E41}
 \end{eqnarray}
 where, $ \mathcal{D}_{\tau} \equiv \partial_{\tau}(\mathfrak{f}(\beta\tau)\partial_{\tau}) $ for some general function, $ \mathfrak{f}(\beta\tau) $.
 
 Let us consider the following Fourier transform for the Green's function in the frequency ($ \mathfrak{w} $) space,
 \begin{eqnarray}
 \mathcal{G}(\tau ,\tau')=\frac{1}{2 \pi}\int d\mathfrak{w}e^{-i\mathfrak{w}\beta(\tau -\tau')}\mathcal{G}(\mathfrak{w}).\label{E42}
 \end{eqnarray}
 
Acting by the L.H.S. of (\ref{E41}), the above equation (\ref{E42}) yields,
\begin{eqnarray}
(\mathcal{D}_{\tau}+\mathfrak{k}(\tau))\mathcal{G}(\tau ,\tau')=-\frac{1}{2 \pi}\int (\mathfrak{w}^{2}\mathfrak{f}(\beta\tau)+i \mathfrak{w}\mathfrak{f}'(\beta\tau)-\mathfrak{k}^{2})e^{-i\mathfrak{w}\beta(\tau -\tau')}\mathcal{G}(\mathfrak{w})d\mathfrak{w}. \label{E43}
\end{eqnarray}
 
Using (\ref{E43}), from (\ref{E41}) we obtain,
\begin{eqnarray}
\mathcal{G}(\mathfrak{w})=\frac{1}{\mathfrak{w}^{2}\mathfrak{f}(\beta\tau)+i \mathfrak{w}\mathfrak{f}'(\beta\tau)-\mathfrak{k}^{2}}
\end{eqnarray}
 which upon substitution in (\ref{E42}) finally yields,
 \begin{eqnarray}
 \mathcal{G}(\tau ,\tau')=\frac{1}{2 \pi}\int d\mathfrak{w}\frac{e^{-i\mathfrak{w}\beta(\tau -\tau')}}{(\mathfrak{w}^{2}\mathfrak{f}(\beta\tau)+i \mathfrak{w}\mathfrak{f}'(\beta\tau)-\mathfrak{k}^{2})}.
 \end{eqnarray}



\begin{thebibliography}{99}

\bibitem{Maldacena:1997re} 
  J.~M.~Maldacena,
  ``The Large N limit of superconformal field theories and supergravity,''
  Int.\ J.\ Theor.\ Phys.\  {\bf 38}, 1113 (1999)
  [Adv.\ Theor.\ Math.\ Phys.\  {\bf 2}, 231 (1998)]
  doi:10.1023/A:1026654312961
  [hep-th/9711200].
  
  \bibitem{Serban:2010sr} 
  D.~Serban,
  ``Integrability and the AdS/CFT correspondence,''
  J.\ Phys.\ A {\bf 44}, 124001 (2011)
  doi:10.1088/1751-8113/44/12/124001
  [arXiv:1003.4214 [hep-th]].
  
  \bibitem{Metsaev:1998it} 
  R.~R.~Metsaev and A.~A.~Tseytlin,
  ``Type IIB superstring action in AdS(5) x S**5 background,''
  Nucl.\ Phys.\ B {\bf 533}, 109 (1998)
  doi:10.1016/S0550-3213(98)00570-7
  [hep-th/9805028].
  
  \bibitem{Bena:2003wd} 
  I.~Bena, J.~Polchinski and R.~Roiban,
  ``Hidden symmetries of the AdS(5) x S**5 superstring,''
  Phys.\ Rev.\ D {\bf 69}, 046002 (2004)
  doi:10.1103/PhysRevD.69.046002
  [hep-th/0305116].
  
  \bibitem{Minahan:2002ve} 
  J.~A.~Minahan and K.~Zarembo,
  ``The Bethe ansatz for N=4 superYang-Mills,''
  JHEP {\bf 0303}, 013 (2003)
  doi:10.1088/1126-6708/2003/03/013
  [hep-th/0212208].
  
  \bibitem{Kruczenski:2003gt} 
  M.~Kruczenski,
  ``Spin chains and string theory,''
  Phys.\ Rev.\ Lett.\  {\bf 93}, 161602 (2004)
  doi:10.1103/PhysRevLett.93.161602
  [hep-th/0311203].
  
  \bibitem{Dimov:2004qv} 
  H.~Dimov and R.~C.~Rashkov,
  ``A Note on spin chain / string duality,''
  Int.\ J.\ Mod.\ Phys.\ A {\bf 20}, 4337 (2005)
  doi:10.1142/S0217751X05020975
  [hep-th/0403121].
  
  \bibitem{Hernandez:2004uw} 
  R.~Hernandez and E.~Lopez,
  ``The SU(3) spin chain sigma model and string theory,''
  JHEP {\bf 0404}, 052 (2004)
  doi:10.1088/1126-6708/2004/04/052
  [hep-th/0403139].
  
  \bibitem{Borsato:2014exa} 
  R.~Borsato, O.~Ohlsson Sax, A.~Sfondrini and B.~Stefanski,
  ``Towards the All-Loop Worldsheet S Matrix for $AdS_3\times S^3\times T^4$,''
  Phys.\ Rev.\ Lett.\  {\bf 113}, no. 13, 131601 (2014)
  doi:10.1103/PhysRevLett.113.131601
  [arXiv:1403.4543 [hep-th]].
  
  \bibitem{Lloyd:2014bsa} 
  T.~Lloyd, O.~Ohlsson Sax, A.~Sfondrini and B.~Stefański, Jr.,
  ``The complete worldsheet S matrix of superstrings on $AdS_3 x S^3 x T^4 $with mixed three-form flux,''
  Nucl.\ Phys.\ B {\bf 891}, 570 (2015)
  doi:10.1016/j.nuclphysb.2014.12.019
  [arXiv:1410.0866 [hep-th]].
  
  \bibitem{Borsato:2016xns} 
  R.~Borsato, O.~Ohlsson Sax, A.~Sfondrini, B.~Stefanski, Jr. and A.~Torrielli,
  ``On the Dressing Factors, Bethe Equations and Yangian Symmetry of Strings on AdS3 x S3 x T4,''
  J.\ Phys.\ A {\bf 50}, no. 2, 024004 (2017)
  doi:10.1088/1751-8121/50/2/024004
  [arXiv:1607.00914 [hep-th]].
  
  \bibitem{Borsato:2013hoa} 
  R.~Borsato, O.~Ohlsson Sax, A.~Sfondrini, B.~Stefanski, Jr. and A.~Torrielli,
  ``Dressing phases of AdS3/CFT2,''
  Phys.\ Rev.\ D {\bf 88}, 066004 (2013)
  doi:10.1103/PhysRevD.88.066004
  [arXiv:1306.2512 [hep-th]].
  
  \bibitem{Borsato:2015mma} 
  R.~Borsato, O.~Ohlsson Sax, A.~Sfondrini and B.~Stefański,
  ``The $\mathrm{AdS}_3\times \mathrm{S}^3\times \mathrm{S}^3\times\mathrm{S}^1$ worldsheet S matrix,''
  J.\ Phys.\ A {\bf 48}, no. 41, 415401 (2015)
  doi:10.1088/1751-8113/48/41/415401
  [arXiv:1506.00218 [hep-th]].
  
  \bibitem{OhlssonSax:2011ms} 
  O.~Ohlsson Sax and B.~Stefanski, Jr.,
  ``Integrability, spin-chains and the AdS3/CFT2 correspondence,''
  JHEP {\bf 1108}, 029 (2011)
  doi:10.1007/JHEP08(2011)029
  [arXiv:1106.2558 [hep-th]].
  
  \bibitem{Babichenko:2009dk} 
  A.~Babichenko, B.~Stefanski, Jr. and K.~Zarembo,
  ``Integrability and the AdS(3)/CFT(2) correspondence,''
  JHEP {\bf 1003}, 058 (2010)
  doi:10.1007/JHEP03(2010)058
  [arXiv:0912.1723 [hep-th]].
  
   \bibitem{Borsato:2016kbm} 
  R.~Borsato, O.~Ohlsson Sax, A.~Sfondrini and B.~Stefański,
  ``On the spectrum of AdS$_3$ × S$^3$ × T$^4$ strings with Ramond–Ramond flux,''
  J.\ Phys.\ A {\bf 49}, no. 41, 41LT03 (2016)
  doi:10.1088/1751-8113/49/41/41LT03
  [arXiv:1605.00518 [hep-th]].
  
  \bibitem{Sax:2012jv} 
  O.~Ohlsson Sax, B.~Stefanski, jr. and A.~Torrielli,
  ``On the massless modes of the AdS3/CFT2 integrable systems,''
  JHEP {\bf 1303}, 109 (2013)
  doi:10.1007/JHEP03(2013)109
  [arXiv:1211.1952 [hep-th]].
    
  
  \bibitem{Stefanski:2004cw} 
  B.~Stefanski, Jr. and A.~A.~Tseytlin,
  ``Large spin limits of AdS/CFT and generalized Landau-Lifshitz equations,''
  JHEP {\bf 0405}, 042 (2004)
  doi:10.1088/1126-6708/2004/05/042
  [hep-th/0404133].
  
  \bibitem{Hofman:2006xt} 
  D.~M.~Hofman and J.~M.~Maldacena,
  ``Giant Magnons,''
  J.\ Phys.\ A {\bf 39}, 13095 (2006)
  doi:10.1088/0305-4470/39/41/S17
  [hep-th/0604135].
  
  \bibitem{Arutyunov:2006gs} 
  G.~Arutyunov, S.~Frolov and M.~Zamaklar,
  ``Finite-size Effects from Giant Magnons,''
  Nucl.\ Phys.\ B {\bf 778}, 1 (2007)
  doi:10.1016/j.nuclphysb.2006.12.026
  [hep-th/0606126].
  
  \bibitem{Kruczenski:2006pk} 
  M.~Kruczenski, J.~Russo and A.~A.~Tseytlin,
  ``Spiky strings and giant magnons on S**5,''
  JHEP {\bf 0610}, 002 (2006)
  doi:10.1088/1126-6708/2006/10/002
  [hep-th/0607044].
  
  \bibitem{Minahan:2006bd} 
  J.~A.~Minahan, A.~Tirziu and A.~A.~Tseytlin,
  ``Infinite spin limit of semiclassical string states,''
  JHEP {\bf 0608}, 049 (2006)
  doi:10.1088/1126-6708/2006/08/049
  [hep-th/0606145].
  
  \bibitem{Spradlin:2006wk} 
  M.~Spradlin and A.~Volovich,
  ``Dressing the Giant Magnon,''
  JHEP {\bf 0610}, 012 (2006)
  doi:10.1088/1126-6708/2006/10/012
  [hep-th/0607009].
  
  \bibitem{Kalousios:2006xy} 
  C.~Kalousios, M.~Spradlin and A.~Volovich,
  ``Dressing the giant magnon II,''
  JHEP {\bf 0703}, 020 (2007)
  doi:10.1088/1126-6708/2007/03/020
  [hep-th/0611033].
  
  \bibitem{Kluson:2007qu} 
  J.~Kluson, R.~R.~Nayak and K.~L.~Panigrahi,
  ``Giant Magnon in NS5-brane Background,''
  JHEP {\bf 0704}, 099 (2007)
  doi:10.1088/1126-6708/2007/04/099
  [hep-th/0703244].
  
  \bibitem{Lee:2008ui} 
  B.~H.~Lee, K.~L.~Panigrahi and C.~Park,
  ``Spiky Strings on AdS(4) x CP**3,''
  JHEP {\bf 0811}, 066 (2008)
  doi:10.1088/1126-6708/2008/11/066
  [arXiv:0807.2559 [hep-th]].
  
  \bibitem{Lee:2008yq} 
  B.~H.~Lee and C.~Park,
  ``Unbounded Multi Magnon and Spike,''
  J.\ Korean Phys.\ Soc.\  {\bf 57}, 30 (2010)
  doi:10.3938/jkps.57.30
  [arXiv:0812.2727 [hep-th]].
  
  \bibitem{Janik:2010gc} 
  R.~A.~Janik, P.~Surowka and A.~Wereszczynski,
  ``On correlation functions of operators dual to classical spinning string states,''
  JHEP {\bf 1005}, 030 (2010)
  doi:10.1007/JHEP05(2010)030
  [arXiv:1002.4613 [hep-th]].
  
  \bibitem{Park:2010vs} 
  C.~Park and B.~H.~Lee,
  ``Correlation functions of magnon and spike,''
  Phys.\ Rev.\ D {\bf 83}, 126004 (2011)
  doi:10.1103/PhysRevD.83.126004
  [arXiv:1012.3293 [hep-th]].
  
  \bibitem{Lee:2011fe} 
  B.~H.~Lee and C.~Park,
  ``Finite size effect on the magnon's correlation functions,''
  Phys.\ Rev.\ D {\bf 84}, 086005 (2011)
  doi:10.1103/PhysRevD.84.086005
  [arXiv:1105.3279 [hep-th]].
  
   \bibitem{Bai:2011su} 
  X.~Bai, B.~H.~Lee and C.~Park,
  ``Correlation function of dyonic strings,''
  Phys.\ Rev.\ D {\bf 84}, 026009 (2011)
  doi:10.1103/PhysRevD.84.026009
  [arXiv:1104.1896 [hep-th]].
  
  \bibitem{Costa:2010rz} 
  M.~S.~Costa, R.~Monteiro, J.~E.~Santos and D.~Zoakos,
  ``On three-point correlation functions in the gauge/gravity duality,''
  JHEP {\bf 1011}, 141 (2010)
  doi:10.1007/JHEP11(2010)141
  [arXiv:1008.1070 [hep-th]].
  
  \bibitem{Hernandez:2011up} 
  R.~Hernandez,
  ``Three-point correlators for giant magnons,''
  JHEP {\bf 1105}, 123 (2011)
  doi:10.1007/JHEP05(2011)123
  [arXiv:1104.1160 [hep-th]].
  
   \bibitem{Ahn:2011zg} 
  C.~Ahn and P.~Bozhilov,
  ``Three-point Correlation functions of Giant magnons with finite size,''
  Phys.\ Lett.\ B {\bf 702}, 286 (2011)
  doi:10.1016/j.physletb.2011.07.011
  [arXiv:1105.3084 [hep-th]].
  
  
  \bibitem{Ahn:2011dq} 
  C.~Ahn and P.~Bozhilov,
  ``Three-point Correlation Function of Giant Magnons in the Lunin-Maldacena background,''
  Phys.\ Rev.\ D {\bf 84}, 126011 (2011)
  doi:10.1103/PhysRevD.84.126011
  [arXiv:1106.5656 [hep-th]].
  
  \bibitem{Bozhilov:2011zp} 
  P.~Bozhilov,
  ``Three-point correlators: Finite-size giant magnons and singlet scalar operators on higher string levels,''
  Nucl.\ Phys.\ B {\bf 855}, 268 (2012)
  doi:10.1016/j.nuclphysb.2011.10.008
  [arXiv:1108.3812 [hep-th]].
  
  \bibitem{Alizadeh:2011yt} 
  D.~Arnaudov and R.~C.~Rashkov,
  ``Three-point correlators: Examples from Lunin-Maldacena background,''
  Phys.\ Rev.\ D {\bf 84}, 086009 (2011)
  doi:10.1103/PhysRevD.84.086009
  [arXiv:1106.4298 [hep-th]].
  
  \bibitem{Minahan:2012fh} 
  J.~A.~Minahan,
  ``Holographic three-point functions for short operators,''
  JHEP {\bf 1207}, 187 (2012)
  doi:10.1007/JHEP07(2012)187
  [arXiv:1206.3129 [hep-th]].
  
  \bibitem{Klose:2011rm} 
  T.~Klose and T.~McLoughlin,
  ``A light-cone approach to three-point functions in $AdS_5 x S^5$,''
  JHEP {\bf 1204}, 080 (2012)
  doi:10.1007/JHEP04(2012)080
  [arXiv:1106.0495 [hep-th]].
  
  \bibitem{Gwak:2012hc} 
  B.~Gwak, B.~H.~Lee and C.~Park,
  ``Correlation functions of the Aharony-Bergman-Jafferis-Maldacena model,''
  Phys.\ Rev.\ D {\bf 87}, no. 8, 086002 (2013)
  doi:10.1103/PhysRevD.87.086002
  [arXiv:1211.5838 [hep-th]].
  
  \bibitem{Georgiou:2013ff} 
  G.~Georgiou, B.~H.~Lee and C.~Park,
  ``Correlators of massive string states with conserved currents,''
  JHEP {\bf 1303}, 167 (2013)
  doi:10.1007/JHEP03(2013)167
  [arXiv:1301.5092 [hep-th]].
  
  \bibitem{Bissi:2011ha} 
  A.~Bissi, T.~Harmark and M.~Orselli,
  ``Holographic 3-Point Function at One Loop,''
  JHEP {\bf 1202}, 133 (2012)
  doi:10.1007/JHEP02(2012)133
  [arXiv:1112.5075 [hep-th]].
  
  \bibitem{Bak:2011yy} 
  D.~Bak, B.~Chen and J.~B.~Wu,
  ``Holographic Correlation Functions for Open Strings and Branes,''
  JHEP {\bf 1106}, 014 (2011)
  doi:10.1007/JHEP06(2011)014
  [arXiv:1103.2024 [hep-th]].
  
  \bibitem{Ryang:2012pm} 
  S.~Ryang,
  ``Three-point correlator of heavy vertex operators for circular winding strings in $AdS_5 \times S^5$,''
  Phys.\ Lett.\ B {\bf 713}, 122 (2012)
  doi:10.1016/j.physletb.2012.05.049
  [arXiv:1204.3688 [hep-th]].
  
  \bibitem{Bozhilov:2015kya} 
  P.~Bozhilov,
  ``Some three-point correlation functions in the $\eta$-deformed AdS$_5 \times S^5$,''
  Int.\ J.\ Mod.\ Phys.\ A {\bf 31}, no. 01, 1550224 (2016)
  doi:10.1142/S0217751X15502243
  [arXiv:1502.00610 [hep-th]].
    
   \bibitem{Bozhilov:2011qf} 
  P.~Bozhilov,
  ``More three-point correlators of giant magnons with finite size,''
  JHEP {\bf 1108}, 121 (2011)
  doi:10.1007/JHEP08(2011)121
  [arXiv:1107.2645 [hep-th]].
  
  \bibitem{Lukowski:2008eq} 
  T.~Lukowski and O.~Ohlsson Sax,
  ``Finite size giant magnons in the SU(2) x SU(2) sector of AdS(4) x CP**3,''
  JHEP {\bf 0812}, 073 (2008)
  doi:10.1088/1126-6708/2008/12/073
  [arXiv:0810.1246 [hep-th]].
  
  \bibitem{Lee:2008sk} 
  B.~H.~Lee, R.~R.~Nayak, K.~L.~Panigrahi and C.~Park,
  ``On the giant magnon and spike solutions for strings on AdS(3) x S**3,''
  JHEP {\bf 0806}, 065 (2008)
  doi:10.1088/1126-6708/2008/06/065
  [arXiv:0804.2923 [hep-th]].
  
   \bibitem{Ciavarella:2010je} 
  A.~Ciavarella and P.~Bowcock,
  ``Boundary Giant Magnons and Giant Gravitons,''
  JHEP {\bf 1009}, 072 (2010)
  doi:10.1007/JHEP09(2010)072
  [arXiv:1007.1674 [hep-th]].
  
  
  \bibitem{David:2008yk} 
  J.~R.~David and B.~Sahoo,
  ``Giant magnons in the D1-D5 system,''
  JHEP {\bf 0807}, 033 (2008)
  doi:10.1088/1126-6708/2008/07/033
  [arXiv:0804.3267 [hep-th]].
  
  
  \bibitem{Arutynov:2014ota} 
  G.~Arutyunov, M.~de Leeuw and S.~J.~van Tongeren,
  ``The exact spectrum and mirror duality of the $(\text{AdS}_5{\times}S^5)_\eta$ superstring,''
  Theor.\ Math.\ Phys.\  {\bf 182}, no. 1, 23 (2015)
  [Teor.\ Mat.\ Fiz.\  {\bf 182}, no. 1, 28 (2014)]
  doi:10.1007/s11232-015-0243-9
  [arXiv:1403.6104 [hep-th]].   
  
  \bibitem{Ali:2015yrs} 
  A.~Mohamed Adam Ali, R.~de Mello Koch, N.~H.~Tahiridimbisoa and A.~Larweh Mahu,
  ``Interacting Double Coset Magnons,''
  Phys.\ Rev.\ D {\bf 93}, no. 6, 065057 (2016)
  doi:10.1103/PhysRevD.93.065057
  [arXiv:1512.05019 [hep-th]].
  
  \bibitem{Beisert:2005tm} 
  N.~Beisert,
  ``The SU(2|2) dynamic S-matrix,''
  Adv.\ Theor.\ Math.\ Phys.\  {\bf 12}, 948 (2008)
  doi:10.4310/ATMP.2008.v12.n5.a1
  [hep-th/0511082].
  
  \bibitem{Dorey:2006dq} 
  N.~Dorey,
  ``Magnon Bound States and the AdS/CFT Correspondence,''
  J.\ Phys.\ A {\bf 39}, 13119 (2006)
  doi:10.1088/0305-4470/39/41/S18
  [hep-th/0604175].
  
  \bibitem{Chen:2006gea} 
  H.~Y.~Chen, N.~Dorey and K.~Okamura,
  ``Dyonic giant magnons,''
  JHEP {\bf 0609}, 024 (2006)
  doi:10.1088/1126-6708/2006/09/024
  [hep-th/0605155].
  
  \bibitem{Bobev:2006fg} 
  N.~P.~Bobev and R.~C.~Rashkov,
  ``Multispin Giant Magnons,''
  Phys.\ Rev.\ D {\bf 74}, 046011 (2006)
  doi:10.1103/PhysRevD.74.046011
  [hep-th/0607018].
  
  \bibitem{Roychowdhury:2016bsv} 
  D.~Roychowdhury,
  ``Multispin magnons on deformed $ AdS_{3}\times S^{3} $,''
  arXiv:1612.06217 [hep-th].
  
  \bibitem{Ahn:2010da} 
  C.~Ahn and P.~Bozhilov,
  ``Finite-Size Dyonic Giant Magnons in TsT-transformed $AdS_5\times S^5$,''
  JHEP {\bf 1007}, 048 (2010)
  doi:10.1007/JHEP07(2010)048
  [arXiv:1005.2508 [hep-th]].
  
  \bibitem{Abbott:2009um} 
  M.~C.~Abbott, I.~Aniceto and O.~Ohlsson Sax,
  ``Dyonic Giant Magnons in $CP^3$: Strings and Curves at Finite $J$,''
  Phys.\ Rev.\ D {\bf 80}, 026005 (2009)
  doi:10.1103/PhysRevD.80.026005
  [arXiv:0903.3365 [hep-th]].
  
  \bibitem{Kalousios:2009mp} 
  C.~Kalousios, M.~Spradlin and A.~Volovich,
  ``Dyonic Giant Magnons on CP**3,''
  JHEP {\bf 0907}, 006 (2009)
  doi:10.1088/1126-6708/2009/07/006
  [arXiv:0902.3179 [hep-th]].
  
   \bibitem{Ryang:2006yq} 
  S.~Ryang,
  ``Three-spin giant magnons in AdS(5) x S**5,''
  JHEP {\bf 0612}, 043 (2006)
  doi:10.1088/1126-6708/2006/12/043
  [hep-th/0610037].
  
  \bibitem{Lunin:2005jy} 
  O.~Lunin and J.~M.~Maldacena,
  ``Deforming field theories with U(1) x U(1) global symmetry and their gravity duals,''
  JHEP {\bf 0505}, 033 (2005)
  doi:10.1088/1126-6708/2005/05/033
  [hep-th/0502086].
  
  \bibitem{Frolov:2005ty} 
  S.~A.~Frolov, R.~Roiban and A.~A.~Tseytlin,
  ``Gauge-string duality for superconformal deformations of N=4 super Yang-Mills theory,''
  JHEP {\bf 0507}, 045 (2005)
  doi:10.1088/1126-6708/2005/07/045
  [hep-th/0503192].
  
  \bibitem{Frolov:2005dj} 
  S.~Frolov,
  ``Lax pair for strings in Lunin-Maldacena background,''
  JHEP {\bf 0505}, 069 (2005)
  doi:10.1088/1126-6708/2005/05/069
  [hep-th/0503201].
  
   \bibitem{Alday:2005ww} 
  L.~F.~Alday, G.~Arutyunov and S.~Frolov,
  ``Green-Schwarz strings in TsT-transformed backgrounds,''
  JHEP {\bf 0606}, 018 (2006)
  doi:10.1088/1126-6708/2006/06/018
  [hep-th/0512253].
  
  \bibitem{Ricci:2007eq} 
  R.~Ricci, A.~A.~Tseytlin and M.~Wolf,
  ``On T-Duality and Integrability for Strings on AdS Backgrounds,''
  JHEP {\bf 0712}, 082 (2007)
  doi:10.1088/1126-6708/2007/12/082
  [arXiv:0711.0707 [hep-th]].
  
  \bibitem{Beisert:2008iq} 
  N.~Beisert, R.~Ricci, A.~A.~Tseytlin and M.~Wolf,
  ``Dual Superconformal Symmetry from AdS(5) x S**5 Superstring Integrability,''
  Phys.\ Rev.\ D {\bf 78}, 126004 (2008)
  doi:10.1103/PhysRevD.78.126004
  [arXiv:0807.3228 [hep-th]].
  
  \bibitem{Hoare:2011wr} 
  B.~Hoare, T.~J.~Hollowood and J.~L.~Miramontes,
  ``q-Deformation of the $AdS_5 x S^5$ Superstring S-matrix and its Relativistic Limit,''
  JHEP {\bf 1203}, 015 (2012)
  doi:10.1007/JHEP03(2012)015
  [arXiv:1112.4485 [hep-th]].
  
  \bibitem{deLeeuw:2011jr} 
  M.~de Leeuw, T.~Matsumoto and V.~Regelskis,
  ``The Bound State S-matrix of the Deformed Hubbard Chain,''
  JHEP {\bf 1204}, 021 (2012)
  doi:10.1007/JHEP04(2012)021
  [arXiv:1109.1410 [math-ph]].
  
  \bibitem{vanTongeren:2013gva} 
  S.~J.~van Tongeren,
  ``Integrability of the ${\rm Ad}{{{\rm S}}_{5}}\times {{{\rm S}}^{5}}$ superstring and its deformations,''
  J.\ Phys.\ A {\bf 47}, 433001 (2014)
  doi:10.1088/1751-8113/47/43/433001
  [arXiv:1310.4854 [hep-th]].
  
  \bibitem{Arutyunov:2012zt} 
  G.~Arutyunov, M.~de Leeuw and S.~J.~van Tongeren,
  ``The Quantum Deformed Mirror TBA I,''
  JHEP {\bf 1210}, 090 (2012)
  doi:10.1007/JHEP10(2012)090
  [arXiv:1208.3478 [hep-th]].
  
  \bibitem{Arutyunov:2012ai} 
  G.~Arutyunov, M.~de Leeuw and S.~J.~van Tongeren,
  ``The Quantum Deformed Mirror TBA II,''
  JHEP {\bf 1302}, 012 (2013)
  doi:10.1007/JHEP02(2013)012
  [arXiv:1210.8185 [hep-th]].
  
  \bibitem{Delduc:2013qra} 
  F.~Delduc, M.~Magro and B.~Vicedo,
  ``An integrable deformation of the $AdS_5 \times S^5$ superstring action,''
  Phys.\ Rev.\ Lett.\  {\bf 112}, no. 5, 051601 (2014)
  doi:10.1103/PhysRevLett.112.051601
  [arXiv:1309.5850 [hep-th]].
  
   \bibitem{Delduc:2014kha} 
  F.~Delduc, M.~Magro and B.~Vicedo,
  ``Derivation of the action and symmetries of the $q$-deformed $AdS_{5} \times S^{5}$ superstring,''
  JHEP {\bf 1410}, 132 (2014)
  doi:10.1007/JHEP10(2014)132
  [arXiv:1406.6286 [hep-th]].
  
  \bibitem{Arutyunov:2013ega} 
  G.~Arutyunov, R.~Borsato and S.~Frolov,
  ``S-matrix for strings on $\eta$-deformed AdS5 x S5,''
  JHEP {\bf 1404}, 002 (2014)
  doi:10.1007/JHEP04(2014)002
  [arXiv:1312.3542 [hep-th]].
  
  \bibitem{Engelund:2014pla} 
  O.~T.~Engelund and R.~Roiban,
  ``On the asymptotic states and the quantum S matrix of the $\eta$-deformed AdS$_{5} \times$ S$^{5}$ superstring,''
  JHEP {\bf 1503}, 168 (2015)
  doi:10.1007/JHEP03(2015)168
  [arXiv:1412.5256 [hep-th]].
  
  \bibitem{Arutyunov:2015qva} 
  G.~Arutyunov, R.~Borsato and S.~Frolov,
  ``Puzzles of $\eta$-deformed AdS$_5 \times$ S$^5$,''
  JHEP {\bf 1512}, 049 (2015)
  doi:10.1007/JHEP12(2015)049
  [arXiv:1507.04239 [hep-th]].
  
  \bibitem{Hoare:2014pna} 
  B.~Hoare, R.~Roiban and A.~A.~Tseytlin,
  ``On deformations of $AdS_n$ x $S^n$ supercosets,''
  JHEP {\bf 1406}, 002 (2014)
  doi:10.1007/JHEP06(2014)002
  [arXiv:1403.5517 [hep-th]].
  
   \bibitem{Lunin:2014tsa} 
  O.~Lunin, R.~Roiban and A.~A.~Tseytlin,
  ``Supergravity backgrounds for deformations of AdS$_{n} \times S^n$ supercoset string models,''
  Nucl.\ Phys.\ B {\bf 891}, 106 (2015)
  doi:10.1016/j.nuclphysb.2014.12.006
  [arXiv:1411.1066 [hep-th]].
  
  \bibitem{Appadu:2015nfa} 
  C.~Appadu and T.~J.~Hollowood,
  ``Beta function of k deformed AdS$_{5}$ × S$^{5}$ string theory,''
  JHEP {\bf 1511}, 095 (2015)
  doi:10.1007/JHEP11(2015)095
  [arXiv:1507.05420 [hep-th]].
  
  \bibitem{Hollowood:2014qma} 
  T.~J.~Hollowood, J.~L.~Miramontes and D.~M.~Schmidtt,
  ``An Integrable Deformation of the $AdS_5 \times S^5$ Superstring,''
  J.\ Phys.\ A {\bf 47}, no. 49, 495402 (2014)
  doi:10.1088/1751-8113/47/49/495402
  [arXiv:1409.1538 [hep-th]].
  
  \bibitem{Hollowood:2015dpa} 
  T.~J.~Hollowood, J.~L.~Miramontes and D.~M.~Schmidtt,
  ``S-Matrices and Quantum Group Symmetry of k-Deformed Sigma Models,''
  J.\ Phys.\ A {\bf 49}, no. 46, 465201 (2016)
  doi:10.1088/1751-8113/49/46/465201
  [arXiv:1506.06601 [hep-th]].
  
  \bibitem{Arutyunov:2014cra} 
  G.~Arutyunov and S.~J.~van Tongeren,
  ``$\mathrm{AdS}_5 \times \mathrm{S}^5$ mirror model as a string sigma model,''
  Phys.\ Rev.\ Lett.\  {\bf 113}, 261605 (2014)
  doi:10.1103/PhysRevLett.113.261605
  [arXiv:1406.2304 [hep-th]].
  
   \bibitem{Kameyama:2014vma} 
  T.~Kameyama and K.~Yoshida,
  ``A new coordinate system for $q$-deformed AdS$_{5} \times$ S$^5$ and classical string solutions,''
  J.\ Phys.\ A {\bf 48}, no. 7, 075401 (2015)
  doi:10.1088/1751-8113/48/7/075401
  [arXiv:1408.2189 [hep-th]].
  
  \bibitem{Kameyama:2014via} 
  T.~Kameyama and K.~Yoshida,
  ``Minimal surfaces in $q$-deformed AdS$_5\times$S$^5$ with Poincare coordinates,''
  J.\ Phys.\ A {\bf 48}, no. 24, 245401 (2015)
  doi:10.1088/1751-8113/48/24/245401
  [arXiv:1410.5544 [hep-th]].  
 
  \bibitem{Khouchen:2014kaa} 
  M.~Khouchen and J.~Kluson,
  ``Giant Magnon on Deformed AdS(3)xS(3),''
  Phys.\ Rev.\ D {\bf 90}, no. 6, 066001 (2014)
  doi:10.1103/PhysRevD.90.066001
  [arXiv:1405.5017 [hep-th]].
  
  \bibitem{Khouchen:2015jfa} 
  M.~Khouchen and J.~Kluson,
  ``D-brane on deformed AdS$_{3} \times$ S$^{3}$,''
  JHEP {\bf 1508}, 046 (2015)
  doi:10.1007/JHEP08(2015)046
  [arXiv:1505.04946 [hep-th]].
  
  \bibitem{Hoare:2014oua} 
  B.~Hoare,
  ``Towards a two-parameter q-deformation of AdS$_3 \times S^3 \times M^4$ superstrings,''
  Nucl.\ Phys.\ B {\bf 891}, 259 (2015)
  doi:10.1016/j.nuclphysb.2014.12.012
  [arXiv:1411.1266 [hep-th]].
  
   \bibitem{Hoare:2015gda} 
  B.~Hoare and A.~A.~Tseytlin,
  ``On integrable deformations of superstring sigma models related to $AdS_n \times S^n$ supercosets,''
  Nucl.\ Phys.\ B {\bf 897}, 448 (2015)
  doi:10.1016/j.nuclphysb.2015.06.001
  [arXiv:1504.07213 [hep-th]].
  
  \bibitem{Arutyunov:2015mqj} 
  G.~Arutyunov, S.~Frolov, B.~Hoare, R.~Roiban and A.~A.~Tseytlin,
  ``Scale invariance of the $\eta$-deformed $AdS_5\times S^5$ superstring, T-duality and modified type II equations,''
  Nucl.\ Phys.\ B {\bf 903}, 262 (2016)
  doi:10.1016/j.nuclphysb.2015.12.012
  [arXiv:1511.05795 [hep-th]].
  
  \bibitem{Sfetsos:2014cea} 
  K.~Sfetsos and D.~C.~Thompson,
  ``Spacetimes for $\lambda$-deformations,''
  JHEP {\bf 1412}, 164 (2014)
  doi:10.1007/JHEP12(2014)164
  [arXiv:1410.1886 [hep-th]].
  
  \bibitem{Demulder:2015lva} 
  S.~Demulder, K.~Sfetsos and D.~C.~Thompson,
  ``Integrable $\lambda$-deformations: Squashing Coset CFTs and $AdS_5\times S^5$,''
  JHEP {\bf 1507}, 019 (2015)
  doi:10.1007/JHEP07(2015)019
  [arXiv:1504.02781 [hep-th]].
  
  \bibitem{Chervonyi:2016ajp} 
  Y.~Chervonyi and O.~Lunin,
  ``Supergravity background of the $\lambda$-deformed AdS$_3 \times$ S$^3$ supercoset,''
  Nucl.\ Phys.\ B {\bf 910}, 685 (2016)
  doi:10.1016/j.nuclphysb.2016.07.023
  [arXiv:1606.00394 [hep-th]].
  
  \bibitem{Chervonyi:2016bfl} 
  Y.~Chervonyi and O.~Lunin,
  ``Generalized $\lambda$-deformations of AdS$_p \times$ S$^p$,''
  Nucl.\ Phys.\ B {\bf 913}, 912 (2016)
  doi:10.1016/j.nuclphysb.2016.10.014
  [arXiv:1608.06641 [hep-th]].
  
  \bibitem{Banerjee:2015nha} 
  A.~Banerjee, S.~Bhattacharya and K.~L.~Panigrahi,
  ``Spiky strings in $\varkappa$-deformed $AdS$,''
  JHEP {\bf 1506}, 057 (2015)
  doi:10.1007/JHEP06(2015)057
  [arXiv:1503.07447 [hep-th]].
  
  
  \bibitem{Banerjee:2016xbb} 
  A.~Banerjee and K.~L.~Panigrahi,
  ``On circular strings in $(AdS_3 \times S^3)_{\varkappa}$,''
  JHEP {\bf 1609}, 061 (2016)
  doi:10.1007/JHEP09(2016)061
  [arXiv:1607.04208 [hep-th]].
  
\bibitem{Kameyama:2015ufa} 
  T.~Kameyama, H.~Kyono, J.~i.~Sakamoto and K.~Yoshida,
  ``Lax pairs on Yang-Baxter deformed backgrounds,''
  JHEP {\bf 1511}, 043 (2015)
  doi:10.1007/JHEP11(2015)043
  [arXiv:1509.00173 [hep-th]].
  
  \bibitem{Panigrahi:2014sia} 
  K.~L.~Panigrahi, P.~M.~Pradhan and M.~Samal,
  ``Pulsating strings on (AdS$_{3}$ × S$^{3}$)$_{ϰ}$,''
  JHEP {\bf 1503}, 010 (2015)
  doi:10.1007/JHEP03(2015)010
  [arXiv:1412.6936 [hep-th]].
  
  \bibitem{Matsumoto:2014ubv} 
  T.~Matsumoto and K.~Yoshida,
  ``Yang-Baxter deformations and string dualities,''
  JHEP {\bf 1503}, 137 (2015)
  doi:10.1007/JHEP03(2015)137
  [arXiv:1412.3658 [hep-th]].
  
  \bibitem{Fateev:1996ea} 
  V.~A.~Fateev,
  ``The sigma model (dual) representation for a two-parameter family of integrable quantum field theories,''
  Nucl.\ Phys.\ B {\bf 473}, 509 (1996).
  doi:10.1016/0550-3213(96)00256-8
  
  
  \end{thebibliography}
\end{document}